\documentclass[a4paper,10pt]{article}
\pdfoutput=1 
\usepackage[utf8]{inputenc}

\usepackage{booktabs}
\usepackage{adjustbox}
\usepackage{multirow}
\usepackage{graphicx}
\usepackage[figuresright]{rotating}
\usepackage[style=numeric-comp, sorting=none]{biblatex}
\usepackage{appendix}

\usepackage{array}
\usepackage{fontenc}
\usepackage{inputenc}
\usepackage{calc}

\usepackage{epstopdf}
\usepackage{lineno}
\usepackage{setspace}
\usepackage{enumitem}
\usepackage{tabularx}
\usepackage{pbox}
\usepackage{caption}
\usepackage{xcolor}

\usepackage{amsmath,amssymb,amsfonts}%
\usepackage{amsthm}%
\usepackage{mathrsfs}%

\usepackage{caption}
\usepackage{subcaption} 
\usepackage{float}      
\usepackage{blindtext}
\usepackage{titlesec}

\usepackage{nicematrix}
\usepackage{dcolumn}
\usepackage{lscape}

\usepackage{authblk}

\usepackage{blindtext}

\newcommand{\PM}{PM$_{2.5}$}
\newcommand{\mugm}{$\mu$g/m$^3$}

\title{\Large Equity in the Distribution of Regulatory \PM\ Monitors}

\author[1,*]{Zoé Haskell-Craig}
\author[2]{Kevin P. Josey}
\author[3]{Patrick L. Kinney}
\author[4]{Priyanka deSouza}

 \affil[1]{\small  Department of Biostatistics, New York University School of Global Public Health, New York, NY}
\affil[2]{\small Department of Biostatistics \& Informatics, Colorado School of Public Health, Aurora, CO}
\affil[3]{\small Department of Environmental Health, Boston University School of Public Health, Boston, MA}
\affil[4]{\small Urban and Regional Planning Department, University of Colorado Denver, Denver, CO}

\affil[*]{\small Email: zjh235@nyu.edu}

\date{} 

\begin{document}

\maketitle

\begin{abstract}

Unequal exposure to air pollution by race and socioeconomic status is well-documented in the U.S. However, there has been relatively little research on inequities in the collection of \PM\ data, creating a critical gap in understanding which neighborhood exposures are represented in these datasets. In this study we use multilevel models with random intercepts by county and state, stratified by urbanicity to investigate the association between six key environmental justice (EJ) attributes (\%AIAN, \%Asian \%Black, \%Hispanic, \%White, \%Poverty) and proximity to the nearest regulatory monitor at the census tract-level across the contiguous 48 states. We also separately stratify our models by EPA region. Our results show that most EJ attributes exhibit weak or statistically insignificant associations with monitor proximity, except in rural areas where higher poverty levels are significantly linked to greater monitor distances ($\beta$ = 0.6, 95\%CI = [0.49, 0.71]). While the US EPA's siting criteria may be effective in ensuring equitable monitor distribution in some contexts, the low density of monitors in rural areas may impact the accuracy of national-level air pollution monitoring.
\end{abstract}

\newpage 




\section{Introduction}
\label{s:intro}

Disparities in \PM\ exposure by race/ethnicity and socioeconomic status (SES) are well-documented \cite{jones_raceethnicity_2014, tessum_pm25_2021, knobel_socioeconomic_2023, nair_environmental_2023, chambliss_local-_2021} and have persisted for over 30 years, despite overall declines in air pollution concentrations \cite{kravitz-wirtz_long-term_2016, colmer_disparities_2020}. There is extensive literature on the disproportionate \PM\ exposure in under-resourced, low-income communities of color compared to wealthier, Whiter neighborhoods. Compounded by increased social stressors and higher baseline disease rates, these inequities exacerbate the health impacts of pollution in these neighborhoods \cite{smith_environmental_2022}. However, relatively little work has focused on inequities in air pollution data collection and monitoring. A key environmental justice (EJ) concern is whether the US Environmental Protection Agency’s (EPA) network of \PM\ monitors is equitably distributed. Monitor placement determines from which communities we have air pollution data (e.g. under-representation of certain neighborhoods in the EPA \PM\ concentration datasets) and the distribution of benefits conferred by the presence of regulatory monitors (e.g. enforcement of regulatory standards; \cite{blacker_measurement_1977}). For instance, counties with monitors that record high pollution levels are subject to restrictions and experienced on average a decrease in \PM\ between $0.47-2.77$ \mugm over a three-year period \cite{sullivan_using_2018}. This study examines the association between EJ attributes at the census tract-level: race/ethnicity, SES, and the location of the nearest regulatory \PM\ monitoring station across the EPA's network. 

The US regulatory air quality monitoring network originated from the Clean Air Act (CAA) of 1970 which established the EPA to develop and maintain national ambient air quality standards (NAAQS) for six widespread pollutants. Monitoring is conducted through two networks: the State or Local Air Monitoring Stations (SLAMS) network, comprised of approximately 1,449 active monitors across all 50 states, Puerto Rico, and the U.S. Virgin Islands, and the Interagency Monitoring of Protected Visual Environments (IMPROVE) network, which operates 159 \PM\ monitors. SLAMS are managed by local government entities (e.g., Florida Department of Environmental Protection, Linn County Health Department, Twenty-Nine Palms Band of Mission Indians), with the number of monitors per county determined by the population size of the Metropolitan Statistical Area (MSA) and recent NAAQS attainment status \cite{40_CFR_AppendixD}. Federal guidelines require \PM\ monitors in urban areas to be sited in locations representative of neighborhood-level air quality (a region with relatively uniform land-use on the scale of 0.5 to 4.0 km$^2$) and adhere to specific distance-to-roadway requirements \cite{40_CFR_AppendixE}. Each state must also maintain at least one regional background and one regional transport site (typically located in sparsely populated areas \cite{40_CFR_AppendixD}). Monitor sites generally fall into one of six location categories. The location is expected to: (i) have the highest concentration of \PM\ in the area covered by the network; (ii) have a typical concentration of \PM\ in densely populated areas; (iii) be near a significant source of \PM\; (iv) represent the general background pollution concentration levels; (v) capture the extent of regional pollutant transport between populated areas; and (vi) where air pollution may impact visibility, damage vegetation, or have other welfare-based impacts \cite{40_CFR_AppendixD}. Monitoring requirements also stipulate that in regions with populations over one million, at least one PM monitor must be co-located at a near-road NO$_2$ station. Furthermore, in regions with extra SLAMS monitors, at least one monitor must be placed in a community at high risk for poor air quality (e.g. near a port) \cite{40_CFR_AppendixD}. The IMPROVE network supplements the SLAMS network by monitoring Class I federally protected lands (e.g. national parks and wilderness areas), with monitors sited away from local pollution sources \cite{noauthor_ucd_2022}. Despite the different siting criteria, both networks provide timely information regarding air pollution, support compliance with air quality standards, and facilitate research into the effects of air pollution \cite{40_CFR_all}.

However, the three goals of the regulatory monitor network each suggest a different siting strategy. For instance, regulating air pollution to meet NAAQS levels requires placing monitors in areas with high \PM\ concentrations, whereas health research may require a network of monitors that represent the typical air pollution exposure of the U.S. population \cite{muller_what_2018}. Although the six categories of locations encompass sites that meet multiple objectives, balancing them is difficult to achieve with the small number of monitors deployed. 

The network of EPA monitors is sparse; less than a third of all U.S. counties contain a monitor, and even within such counties, the network is too limited to accurately capture the variability in pollution that often exists at the census tract or block-group area level \cite{considine_investigating_2023, english_imperial_2017, stuart_social_2009}. As such, the data recorded from these monitors may be unrepresentative of the pollution levels experienced by many residents in these counties. There is also some evidence that industry emitters game the system. For instance, one study demonstrated that air quality is worse on days during which regulatory monitors are not collecting data \cite{zou_unwatched_2021}. Furthermore, research has shown that regulatory monitor data may be particularly unrepresentative of the exposures faced by individuals of color, and those living in poverty \cite{stuart_social_2009}. \cite{stuart_social_2009} conducted a case study on all EPA air quality system (AQS) monitors in Tampa, Florida, finding that individuals who are Black, Hispanic, and those living in poverty disproportionately lived closer to sources of air pollution and further from monitoring sites than the overall county population. The measurement bias induced by regulatory monitor siting may be the result of local agencies trying to decrease discovery of pollution ``hotspots" by strategically siting monitors away from sources of pollution, which are over-represented in low-income communities and communities of color \cite{lee_prioritizing_2020,cushing_historical_2023}. Grainger and Schreiber (2019) find that counties slated to receive a new NO$_2$ monitor are more likely to site monitors away from pollution hotspots, except for hotspots in neighborhoods that are Whiter and wealthier \cite{grainger_discrimination_2019}. Hotspot avoidance may ultimately contribute to documented disparities in misclassifications of NAAQS attainment across racial and ethnic groups \cite{sullivan_using_2018}.

Unequal monitoring may perpetuate disparities in exposure misclassification from models developed from this dataset. EPA regulatory monitors comprise the largest network of high-quality ground-based \PM\ data that are routinely used for a variety of purposes, including as input for chemical transport models (e.g. \cite{di_hybrid_2016}), as the ground-based measurements for developing \PM\ measurements from remote sensing-derived aerosol optical depth (AOD) \cite{van_donkelaar_monthly_2021, shaddick_global_2021}, to assess the accuracy \cite{feenstra_performance_2019} and calibration of low-cost sensors \cite{datta_statistical_2020}, and as the input dataset for empirical (``land-use regression") models -- including for the six major models producing national estimates \cite{bechle_intercomparison_2023}. As Bechle et al. (2023)  \cite{bechle_intercomparison_2023} state ``whatever strengths or weaknesses exist in using EPA monitors (and their locations) for empirical models, those likely impact all of the [empirical] models." Measurements from the network and exposure assessments from these models are regularly used in health research assessing the impact of \PM\ on cardiovascular disease \cite{chi_individual_2016}, low birth weight \cite{yang_impact_2017}, and diabetes among others \cite{coogan_pm25_2015, legro_effect_2010, magzamen_differential_2021, parker_air_2009, roberts_association_2014, vopham_ambient_2018}.

Despite the critical role of regulatory \PM\ monitors in pollutant data collection and monitoring, little work has investigated whether there are systematic racial/ethnic or income disparities in the location of these monitors. Previous work has focused on whether such monitors robustly capture pollution variability \cite{kelp_new_2022} or hotspots \cite{wang_us_2024}. To our knowledge, only three studies have assessed the network at a national level from an environmental justice perspective \cite{grainger_discrimination_2019,miranda_making_2011, pedde_representativeness_2024}. Grainger and Schreiber (2019) \cite{grainger_discrimination_2019} focused on the placement of new NO$_2$ monitors, rather than the existing monitor network as a whole. Miranda et al.  (2011) \cite{miranda_making_2011} assessed the presence/absence of \PM\ and O$_3$ monitors at a county level, without considering their location with respect to sub-county variability in socioeconomic and demographic characteristics. Pedde and Adar (2024) \cite{pedde_representativeness_2024} addressed concerns with the use of regulatory data as input in empirical models by investigating the representativeness of monitor locations with respect to covariates typically included as predictors in these models (e.g. population density, land-use, etc.). In particular, they stratify by race and calculate the fraction of the US population that lives outside (above/below the max/min value) of the numeric range of covariate values. Other work on disparities in monitor placement has focused on low-cost sensors, such as PurpleAir monitors, rather than the more costly monitors that the EPA deploys \cite{desouza_distribution_2021, mullen_exploring_2022, sun_socioeconomic_2022}. Our goal is to investigate whether there are disparities by race/ethnicity and SES in census tract distance to the nearest \PM\ monitor across the SLAM and IMPROVES networks.

\section{Methods}
\label{s:methods}

\subsection{Data}
Regulatory \PM\ monitor locations were obtained from the US EPA's Outdoor Air Quality dataset of annual pollutant concentrations for 2022 \cite{us_environmental_protection_agency_air_2023}. For this analysis, we included only monitors recording local \PM\ conditions, resulting in a final sample size of $969$ monitors. For each of the $84{,}050$ census tracts in the contiguous US (excluding Alaska and Hawaii), we computed the distance from the 2020 population-weighted centroid, obtained from the IPUMS National Historical Geographic Information System \cite{manson_2020_2023}, to (a) the nearest monitor site and (b) the nearest monitor site located within the same county. For the latter, we excluded tracts located in counties without a monitoring site. We use distance to the nearest monitor as a proxy for monitor coverage for the following reasons. Simply considering the tract within which a monitor is located (``unit-hazard coincidence" approach) ignores neighboring tracts when monitors are located near the boundary \cite{bullard_toxic_2008}. Assigning monitors to tracts if they fall within a specific radius (``buffer" approach) requires specifying a buffer size, which may not reflect differing drivers of pollution and pollutant transport conditions across the country. 

Considering axes of EJ widely-used in the literature \cite{casey_measuring_2024, nunez_environmental_2024, gonzalez_temporal_2023}, we analyzed whether there exist disparities in neighborhood proximity to regulatory \PM\ monitors by racial composition (\%non-Hispanic White, \%Black, \%Hispanic, \%Asian, and \%American Indian or Alaskan Native (AIAN)) and socioeconomic status (operationalized as the proportion of the population living below the federal poverty line). We obtained census-tract level demographic and socioeconomic data, including median household income, proportion of population living below the federal poverty line, and race/ethnicity in 2020 (the most recent data available) from the American Community Survey (ACS), downloaded from the Social Determinants of Health database \cite{agency_for_healthcare_research_and_quality_social_2023}.  Census tracts with missing demographic data were excluded from the analysis, as were tracts with a total population of less than 100 to avoid noise associated with small population size ($2{,}937$ tracts, $3.4\%$), resulting in a final sample size of $82{,}329$ tracts. For the 2020 census, the US Census Bureau defined the tracts as urban if they were within a territory that contained a densely settled core with a minimum of $2{,}000$ housing units or a population of $5{,}000$. A list of urban census tracts from the 2020 decennial census is available from the Census website \cite{us_census_bureau_block-level_nodate}. As federal guidelines and processes for selecting monitor sites differ between urban and rural counties (e.g. in rural areas monitoring locations are chosen to represent capture long-range transport) all analysis were stratified by urbanicity.

2022 annual average \PM\ concentrations for each census tract were obtained from the Atmospheric Analysis Composition Group's satellite-derived \PM\ dataset, created using a statistical fusion of satellite aerosol optical depth, GEOS-Chem simulation of emissions, and information from ground monitors \cite{van_donkelaar_monthly_2021}. As satellite-derived \PM\ data fall on a raster grid at a $0.01^{\circ} \times 0.01^{\circ}$ resolution, we computed the spatially weighted mean \PM\ for grid cells that overlap each census tract. To compare areas with similar high or low pollution levels and to reduce the impact of outliers, we computed the relative \PM\ Z-score for each census tract. That is, we denoted the \PM\ exposure for each tract as the number of standard deviations (SDs) from the overall mean. It should be noted that the census tract geographic boundaries did not change between 2020 (the most recent year of the ACS) and 2022, except for Connecticut, where we used the 2020 tract boundaries.

\subsection{Statistical Analysis}

We apply two approaches to assess our research questions: (1) a multilevel model that allows for random effects at the county and state-level and a (2) a Bayesian multilevel model that also accounts for spatial structure in the data. We present the methodology and results for the non-spatial model here, as the model is simpler to interpret and the results are very similar from the two approaches. The method and results for the Bayesian spatial multilevel model is detailed in Appendix \ref{app:spatial}. 

\subsubsection{Apportionment of variance in monitor proximity}

To determine the proportion of variance in monitor proximity occurring at different spatial scales (within-county, between counties within each state, between states) we fit four multilevel linear models with random intercepts at the county and state level. Including only census tracts in counties with at least one \PM\ monitor, and considering proximity only to monitors located within the same county as the census tract, we fit:(i) a null model; (ii) a model adjusting for urbanicity and population. For all census tracts in the sample, considering their proximity to the nearest monitor regardless of whether this monitor was collocated within the same county or state we fit: (iii) a null model; and (iv) an adjusted model. Following the approach used by Boing et al. in \cite{boing_air_2022} and \cite{boing_quantifying_2020}, we calculated the proportion of variance explained at the different spatial levels by dividing the variance at each level by the sum of observed variance across all levels. 

\subsubsection{Association between census tract EJ attributes and monitor proximity}

Our goal is to identify whether there are systematic associations between the EJ attributes of a census tract and the proximity to the nearest monitor. Any such model should also take into account geographic differences (size, shape, location) of each state and account for monitoring requirements that differ at a county level. As the EPA subdivides the US into ten regions with corresponding regional offices responsible for the execution of programs within those states, we first fit models for the US as a whole, and then stratified by EPA region.

To assess the association between EJ attributes and proximity to regulatory \PM\ monitors across the US we fit a series of multilevel models with random intercepts at the county and state level (eq. \ref{eq:MLM}). Also termed 'hierarchical' \cite{gelman_bayesian_2013} or 'mixed-effect' \cite{fong_bayesian_2010} models, these are defined by the inclusion of a normally distributed random intercept (also called a 'random effect') to capture variation between groups of observations that are clustered - in our case, census tracts are clustered within counties which in turn are clustered within states \cite{garson_hierarchical_2013}. Following the literature, we fit one model for each EJ attribute under consideration (6 models for the main analysis; \cite{casey_measuring_2024,nunez_environmental_2024,cushing_toxic_2023}). We consider several racial/ethnic identities because systemic racism may occur differently for different racial categories, and fit separate models for each attribute to avoid over-controlling by multiple racial/ethnic groups simultaneously. As monitoring requirements are partially informed by population and pollution levels, we controlled for population density and \PM\ concentrations. In models with percent Black, Hispanic, Asian, or AIAN as the main predictor we also controlled for SES and the proportion of the population identifying as White since proportion White may or may not co-vary with Black, Hispanic, Asian, or AIAN populations and SES \cite{altschuler_local_2004, estabrooks_resources_2003, haeberle_good_1986, moore_availability_2008, nunez_environmental_2024}. In models where SES is the main predictor we controlled for the proportion of non-White residents for the same reason. Equation \ref{eq:MLM} displays the model specification for $Y_{ijk}$, the log-distance to the nearest monitor of census tract $i$ in county $j$ and state $k$, as a linear function of the intercept, $\beta_0$, the EJ attribute as the main predictor, a vector of additional covariates $\vec{X}$, with a random intercept at the county, $\gamma_{jk}$, and state, $\alpha_k$, level (See Appendix B section \ref{sec:main_models} for model specifications including all covariates). As a sensitivity analysis, we re-fit the same models with SES operationalized as median household income (Appendix B section \ref{sec:sensitivity_models}). All statistical analyses were conducted in R version 4.4.0 \cite{r_core_team_r_2014}. 

\begin{linenomath}
\begin{equation}\label{eq:MLM}
Y_{ijk} = \beta_{0} + \beta_{1} EJ + \vec{\beta_{x}}\vec{X} + \gamma_{jk} + \alpha_{k} + \epsilon_{ijk}
\end{equation}
\end{linenomath}

\vspace{-2em}

\section{Results}

We obtained a final sample size of $82{,}329$ census tracts across the 48 states: $70{,}483$ urban and $11{,}846$ rural. Table \ref{tab:descriptives} displays descriptive statistics by urbanicity. We found the average distance between the population-weighted centroid of each census tract and the nearest \PM\ monitor active on at least one day of 2022 to be $23.5$ km (sd = $28.3$ km). The distance varies by urbanicity, with monitors located an average of $18.8$ km (sd = $24.2$ km)  and $51.9$ km (sd = $34.2$ km) away in urban and rural census tracts, respectively. We further restricted the data to counties with at least one monitor, and calculated the distance to the nearest monitor co-located within the same county as the census tract. Out of the $3{,}107$ counties included in our analysis, $612$ contained at least one monitor, for a total of $55{,}661$ census tracts ($68\%$). For these counties, monitors were located on average $12.8$ km (sd = $71.6$ km) from urban and $34.4$ km (sd = $137.9$ km) from rural census tracts. $6{,}170$ ($12\%$) of the $53{,}052$ urban census tracts and $396$ ($15\%$) of the $2{,}609$ rural tracts located in a county with at least one monitor are nearer to a monitor in a neighboring county than their own. 

\begin{table}[!ht]
    \centering
    
    \begin{tabular}{l c c c}
    
          & \multicolumn{3}{c}{Mean (sd)}  \\ \cline{2-4}
          \multirow[c]{3}{*}{} & Urban & Rural & All \\ 
          & (n = 70483) & (n = 11846) & (n = 82329)  \\ \hline
        \multicolumn{1}{l|}{Distance to nearest monitor (km)} & & &   \\ 
        \multicolumn{1}{r|}{All tracts} & 18.8 (24.2) & 51.9 (34.2) & 23.5 (28.3)  \\ 
        \multicolumn{1}{r|}{Tracts in county with } & & & \\
        \multicolumn{1}{r|}{monitor (n = 55661)} & 12.8 (71.6) & 34.4 (137.9) & 13.8 (76.1) \\ \hline
        \multicolumn{1}{l|}{EJ attributes} & & & \\ 
        \multicolumn{1}{r|}{\% AIAN} & 0.6 (2) & 2 (10) & 0.9 (4)  \\ 
        \multicolumn{1}{r|}{\% Asian} & 6 (10) & 0.6 (1.3) & 5 (9) \\ 
        \multicolumn{1}{r|}{\% Black}& 15 (22) & 7 (15) & 13 (21)  \\ 
        \multicolumn{1}{r|}{\% Hispanic} & 19 (22) & 7 (13) & 17 (22) \\ 
        \multicolumn{1}{r|}{\% White} & 58 (30) & 82 (21) & 61 (30)  \\ 
        \multicolumn{1}{r|}{\% Below poverty} & 13 (11) & 14 (9) & 13 (11) \\ 
        \multicolumn{1}{r|}{Median HH income (100k)} & 0.72 (0.37) & 0.57 (0.19) & 0.70 (0.35) \\ \hline
    \end{tabular}
    \caption{Monitor proximity and EJ attribute descriptive statistics. For each census tract we calculated the distance from the population-weighted centroid of the tract to the nearest regulatory monitor. We also subset the data to tracts in counties containing at least one monitor. We stratified all analyses by urbanicity. The mean and standard deviation (sd) of census tract EJ attributes are displayed.}
     \label{tab:descriptives}
\end{table}

\begin{figure}
\includegraphics[width=1\linewidth]{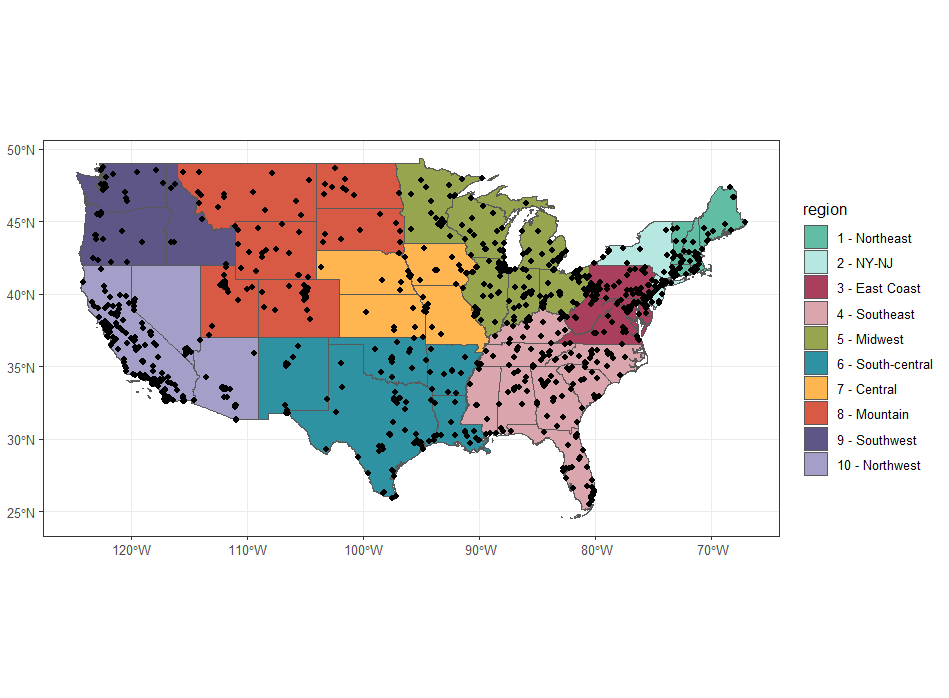}
\caption{Map displaying the location of EPA regulatory \PM\ monitors (black triangles) in each state. States are grouped by EPA regions. Monitors are sparse across the US: only $612$ out of the $3{,}107$ counties contain at least one monitor and monitors are located on average $18.8$ km and $51.9$ km from urban and rural census tracts respectively.}
\label{fig:map} 
\end{figure}

\begin{figure}
\includegraphics[width=1\linewidth]{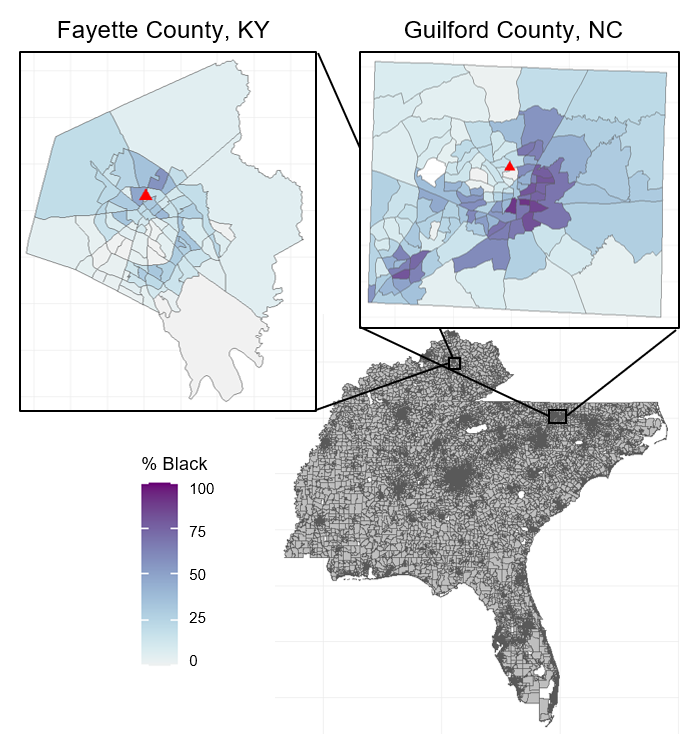}
\caption{ Map of census tracts in the Southeast EPA region. Census tracts with the percent
of the population identifying as Black are displayed in blue for Lexington in Fayette County, KY and Greensboro in Guilford County, NC. These cities each have a similar population size ($320{,}000$ and $300{,}000$ respectively) and one regulatory \PM\ monitor (denoted by the red triangles). However, in Fayette County, the \PM\ monitor is located in and near census tracts with a larger Black population, while in Guilford County the \PM\ monitor is located in and near census tracts where a smaller proportion of the population is Black. With just one monitor per county, the association between EJ attribute and proximity may vary widely between counties.}
\label{fig:sparsity}
\end{figure}

The location of \PM\ monitors across the contiguous US are displayed in figure \ref{fig:map}. The sparsity of monitoring stations - most counties contain a single monitor - results in large variability in the distance to the closest monitor, and the association between EJ attributes and this distance across counties. For example, consider two counties in EPA region 4: Fayette County, KY with a population of approximately $320{,}000$, encompassing the city of Lexington (pop. $320{,}000$) and Guilford County, NC with a population of roughly $550{,}000$ where Greensboro, NC is located (pop. $300{,}000$ thousand) (Figure \ref{fig:sparsity}). Each county operates exactly one \PM\ monitor, located within the city. Focusing on a single EJ attribute -- the proportion of the census tract that identifies as Black --we see that in Fayette County, the \PM\ monitor is located in and near census tracts with a large Black population, whereas in Guilford County the opposite is true.

Table \ref{tab:variance} displays the proportion of the total variance by state and county, before and after adjusting for census tract urbanicity and population size. Across the contiguous US as a whole, the largest proportion of variance in distance to monitors ($52\%$) can be attributed to the county-level, proceeding from the relatively small number of counties required to maintain regulatory monitors due to population size or selected for background monitoring/regional transportation sites. Within-county, that is, between-census tract residual differences remain an important source of variability ($31\%$). When restricting the sample to census tracts located within a county with at least one monitor, the largest proportion of variability is attributed to within-county differences ($67\%$), followed by between-county differences ($27\%$). Controlling for census population and urbanicity does not greatly change the apportionment of variance (difference $\le 4\%$).

\begin{table}[H] 

\newcolumntype{C}{>{\centering\arraybackslash}X}
\begin{tabularx}{\textwidth}{C|CCCC}
\toprule
\textbf{Level}	& Model i \textsuperscript{1}	& Model ii\textsuperscript{1}\textsuperscript{2} &  Model iii	& Model iv\textsuperscript{2} \\
\midrule
State	  & 0.05			& 0.05         &   0.16    & 0.15\\
County	  & 0.27			& 0.24         &  0.52    & 0.51 \\
Residual  & 0.67			& 0.71        &    0.31    & 0.34\\
\bottomrule
\end{tabularx}
\noindent{\footnotesize{\textsuperscript{1} Models (i) and (ii) include only census tracts in counties containing at least one monitor} \\
\textsuperscript{2} Models (ii) and (iv) adjust for log(population) and urbanicity}
\caption{Proportion of variance attributed to each spatial scale. Variance is calculated by fitting a null multilevel model for the distance to the nearest monitor for models (i) and (iii), and a model controlling for population and urbanicity for models (ii) and (iv). We restrict the data to census tracts located in counties with at least one monitor for models (i) and (ii). }
 \label{tab:variance}
\end{table}

\begin{figure}[H]
\includegraphics[width=1.3\linewidth]{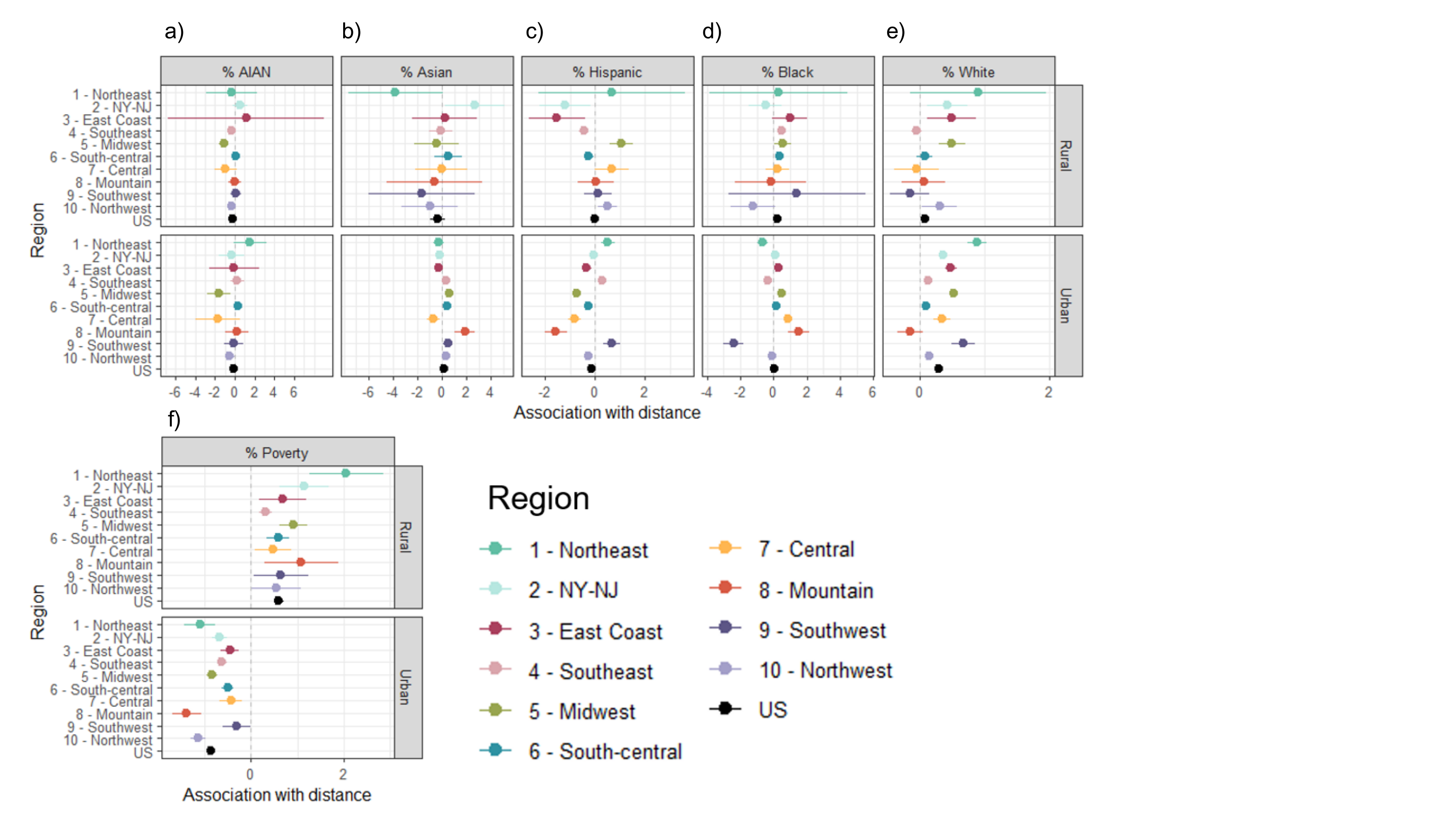}
\caption{ Coefficients on the association between census tract EJ attribute and distance to the nearest monitor for (a) \%AIAN (b) \%Asian (c) \%Hispanic (d) \%Black (e) \%White (f) \% living below the poverty line. The circles represent coefficient point estimate and the bars the 95\% CI. Results from the separate models run for each US EPA region are colored by region, the black point and line represent the estimate from the model for the US as a whole. For the most part, the relationship between EJ attribute and monitor proximity is weak (point estimate near zero) or insignificant (95\% CI bars cross $0$) for the US as a whole. The exception is for \% poverty: for rural census tracts a higher percent of the population living below the poverty line is significantly associated with a farther distance to the nearest monitor for all regions and the US as a whole.}
\label{fig:MLMcoeffs}
\end{figure}

The point estimate and $95\%$ confidence interval for the coefficients of interest in each multilevel model are presented in figure \ref{fig:MLMcoeffs}. Positive coefficients indicate that the attribute is associated with a larger distance to the nearest \PM\ monitor, while negative coefficients indicate a shorter distance. Among rural census tracts in the US-wide model there is a small association between a higher proportion of the population identifying as AIAN and a shorter distance to the nearest monitor ($\beta = -0.19$, $95\%$ CI $= [-0.31, -0.08]$) and a higher proportion of the population identifying as Black with a longer distance ($\beta = 0.25$, $95\%$ CI $= [0.14, 0.36]$) (Figure \ref{fig:MLMcoeffs}, US-wide model results in black). The coefficient of poverty is positive and relatively large in magnitude ($\beta = 0.60$, $95\%$ CI $= [0.50, 0.70]$), indicating that having a higher proportion of the population living below the poverty line in a rural census tract is associated with a further distance to the nearest monitor. For instance, for rural tracts, each $10\%$ increase in poverty is associated with $6\%$ increase in distance on average, all else equal (all coefficient estimates presented in Appendix B tables \ref{tb:mlm_rural}). For rural tracts, all other EJ attributes model coefficients are weak (near-zero, $|\beta| < 0.1$) or non-statistically significant (confidence intervals spanning zero). 

Conversely, for urban census tracts higher poverty is associated with a shorter distance to the nearest monitor ($\beta = -0.88$, $95\%$ CI $= [-0.93, -0.83]$). The coefficients for proportion White and Asian are small in magnitude and positive ($\beta = 0.30$, $95\%$ CI $= [0.27, 0.33]$ and $\beta = 0.17$, $95\%$ CI $= [0.11, 0.23]$ respectively), and the coefficient on Hispanic is small in magnitude and negative ($\beta = -0.12$, $95\%$ CI $= [-0.16, -0.08]$). All other EJ attributes have weak or non-statistically significant associations. In a sensitivity analysis, the magnitude and sign of coefficients remain consistent when substituting the proportion of the census tract living below the poverty line for median household income (Appendix B table \ref{tb:mlmS_rural} and figure \ref{fig:sensitivity}).

 Stratifying by regions, we find some heterogeneity in the coefficients on the EJ attributes (figure \ref{fig:MLMcoeffs}). Specifically, for \%Hispanic in rural tracts the US-wide model shows no statistically significant association overall, yet stratifying by region we find a significant positive association for rural tracts in the Midwest ($\beta = 1.06$, $95\%$ CI = $[0.60, 1.55]$) and negative association for tracts in the NY-NJ region ($\beta = -1.18$, $95\%$ CI = $[-2.20, -0.16]$), the East Coast ($\beta = -1.52$, $95\%$ CI = $[-2.64, -0.40]$) and the Southeast ($\beta = -0.44$, $95\%$ CI = $[-0.64, -0.24]$). Similarly, we find a positive association between \%Hispanic and monitor proximity for urban tracts in the Northeast ($\beta = 0.52$, 95\% CI = $[0.23, 0.80]$) and Southwest ($\beta = 0.69$, 95\% CI = $[0.35, 1.03]$), and a negative association for tracts in the Midwest ($\beta = -0.71$, 95\% CI = $[-0.81, -0.61]$), Central ($\beta = -0.80$, 95\% CI = $[-1.07, -0.54]$), and Mountain regions ($\beta = -1.56$, 95\% CI = $[-1.99, -1.12]$). We also observe some regional differences in the association between monitor proximity and \%Black and \%Asian for urban census tracts (Appendix B tables \ref{tb:coeffs_region}). For other EJ attributes the regional sub-analyses are consistent with the results of the US-wide model. The coefficient on \%AIAN is small and negative (Midwest) or with a confidence interval spanning zero for all regional models; the coefficient on \%White is small and positive or zero for all regional models. 

Examining the association between census tract poverty and \PM\ regulatory monitoring, we observe a consistent and relatively strong association across all regions, with a positive association for rural and negative association for urban census tracts. The magnitude of the association is particularly large for rural tracts in the Northeast ($\beta = 2.06$, 95\% CI = $[1.26, 2.86]$) and urban tracts in the Northeast ($\beta = -1.11$, 95\% CI = $[-1.44, -0.78]$), Mountain ($\beta = -1.40$, 95\% CI = $[-1.71, -1.09]$), and Northwest ($\beta = -1.15$, 95\% CI = $[-1.32, -0.98]$) regions.  

A comparison of point estimates and 95\% credible intervals shows a high degree of similarity between the estimate of associations between monitor proximity and census tract EJ attribute between the Bayesian models with and without a spatial component, and the frequentist multilevel model (Appendix A figure \ref{fig:INLA_comparison}).

\section{Discussion}
\label{s:discuss}

Consistent with the literature, we find that the network of regulatory \PM\ monitors is sparse, with the nearest monitor located $18.8$ km and $51.9$ km away from urban and rural census tracts respectively. Consequently, the pollution levels experienced in many census tracts may not be captured by regulatory monitors. Although generally high or low concentrations of fine particles such as \PM\ are considered to be a regional (e.g. state) feature \cite{wilson_fine_1997}, \PM\ concentrations do vary substantially within cities \cite{pinto_spatial_2004} and even within neighborhoods \cite{kinney_airborne_2000}. Given the large within-county variance in proximity to monitors, some counties and census tracts may benefit from regulatory monitoring while others may not. Furthermore, monitor sparsity results in high heterogeneity in the association between monitor proximity and census tract EJ attributes across the country. Since census tracts are often clustered by race/ethnicity and income, randomly allocating a single monitor within a county will likely result in disparities locally - although not globally. This is not to say that monitor siting decisions within any given county or state are or are not driven by neighborhood socioeconomic and demographic composition - but that this cannot be determined from the location of the \PM\ monitors alone.

We do not find a systematic disparity by race/ethnicity or SES in proximity to the nearest \PM\ monitor across the US, with the exception of rural poverty. This aligns with previous work by Pedde and Adar (2024) \cite{pedde_representativeness_2024}, who conclude that there are little to no racial or ethnic differences in the representativeness of the covariate distribution surrounding \PM\ monitors at a national level. Similar to \cite{pedde_representativeness_2024}, we also detect variability in the association between EJ attributes and monitor proximity across subregions of the US.  This is the first study of its kind to define equity in monitoring at the neighborhood level by proximity. Related work by \cite{grainger_discrimination_2019} found a statistically significantly higher probability of a new NO$_2$ monitor being located within a neighborhood with a higher proportion of White residents. However, this result only held for areas with low concentrations of NO$_2$, and did not test whether the difference in probability of new monitor placement is strong enough to affect the overall network of monitors. Our results demonstrate that the dataset produced by the EPA's regulatory \PM\ monitor network does not undersample \PM\ concentrations from neighborhoods with EJ attributes. Algorithms for estimating \PM\ exposure that utilize this dataset are unlikely to suffer from sampling bias by race/ethnicity or SES \cite{ravishankar_provable_2023}. This suggests that the US EPA siting criteria guidelines may successfully be adopted in other contexts to produce an equitable distribution of monitors, for instance in allocating low-cost air quality monitors. 

However, additional attention should be paid to monitoring in areas with rural poverty. The low density of monitors in rural areas has already been noted in the literature as impacting the accuracy of national-level air pollution monitoring \cite{bechle_intercomparison_2023}. \cite{bechle_intercomparison_2023} demonstrate that models vary more widely in their estimates of pollution in rural areas, casting doubt on the accuracy of at least some of those models (as opposed to urban areas, where models largely agree). Although \PM\ concentrations are generally lower in rural as opposed to urban areas, disparities in exposure by race/ethnicity still exist \cite{tessum_pm25_2021, nair_environmental_2023}. However, little previous research has considered disparities in exposure by SES in rural areas. For instance, both \cite{nair_environmental_2023} and \cite{tessum_pm25_2021} find disparities in \PM by race / ethnicity at a range of urbanization levels and by SES, yet do not explore the interaction of income and urbanicity. Although focusing on urban areas, \cite{knobel_socioeconomic_2023} determined that agriculture, dust, metal processing industry, and vehicle traffic sources are associated with \PM\ disparities by income; other research has determined that traffic exposure disparities by income are largely driven by emissions from interstate highways \cite{lee_prioritizing_2020}. These may be important drivers of \PM\ exposure in low-income rural areas that are not currently surveyed by regulatory monitors. 

Some limitations of the present study include the restriction of the analysis to the contiguous US (without Alaska, Hawaii, or Puerto Rico for instance). As well, there may be unmeasured covariates that are associated with monitor site selection. We did not include information on local decision-making processes or local criteria for monitor siting, beyond broad federal guidelines. Furthermore, we assume that distance to the nearest monitor can be used as a proxy to measure how well monitors capture \PM\ emissions from a given census tract. This includes the implicit assumption that the linear distance determines the correlation between levels of \PM\ in the census tract and at the monitor, as we do not consider geographic features (e.g. mountains/ravines) or pollution transport patterns. Efforts to reduce pollution at a specific monitoring location might not necessarily lead to a decrease in pollution levels in the nearby census tracts. Future work should examine the extent to which neighborhoods benefit from hosting regulatory monitors. In addition, researchers should investigate the distribution of monitors for other pollutants that are important drivers of air quality. 

Despite these limitations, this is the first study to examine whether there exist systematic disparities in the location of regulatory \PM\ monitors across the US as a whole by race/ethnicity and socioeconomic status. We include tracts located both in counties with and without regulatory monitors and consider the spatial nature of the data - including the clustering of tracts by race/ethnicity and SES and by proximity to monitors. In general, we find that the network of regulatory \PM\ monitors is equitably distributed, and pollution concentrations experienced by neighborhoods with EJ attributes are captured in the resulting dataset. However, in rural areas, census tracts with lower SES are located further, on average, from monitors than higher SES tracts after controlling for \PM\ levels. Access to regulatory monitoring can be considered a key component of environmental justice. As such, efforts should be made to improve monitoring in low-income rural areas.


\printbibliography
\newpage

\appendix
\renewcommand{\thefigure}{S\arabic{figure}} 
\renewcommand{\thetable}{S\arabic{figure}}
\renewcommand{\thesection}{Appendix \Alph{section}}

\section{Bayesian multilevel model with spatial error}\label{app:spatial}




\renewcommand{\thefigure}{S\arabic{figure}} 
\renewcommand{\thetable}{S\arabic{figure}}
\renewcommand{\thesection}{\Alph{section}}


Given the spatial nature of the data (census tracts are often clustered by race/ethnicity and SES \cite{arcaya_multi-level_2018, cooke_geographic_1999, dwyer_contained_2012} and neighboring tracts have a similar proximity to monitors), we are concerned that this structure may violate the underlying assumptions of the multilevel modelling approach employed. In particular, the assumption that the residuals are independent and identically distributed (i.i.d.), follow a normal distribution with a constant variance, and are independent of spatial location. Spatial data, however, rarely adheres to these assumptions, and model estimates may be biased and/or may underestimate the width of confidence intervals as they incorrectly treat each observation as independent \cite{de_graaff_general_2001} \cite{wall_close_2004} \cite{congdon_applied_2014}. As such, we additionally check the spatial independence of the multilevel model residuals (section \ref{sec:moran}) and, if we detect spatial autocorrelation, fit a spatial model (section \ref{sec:INLA}).

\subsection{Methods}

\subsubsection{Defining neighboring census tracts}\label{sec:neighboring}

To conduct any spatial analyses, we first must define which units are nearby in space (neighbors). All analysis were stratified by census tract urbanicity, resulting in two disjoint sets of census tracts with the following spatial patterns: (i) a set with ``islands" of urban tracts, predominantly clustered in urban centers and non-contiguous, and (ii) a set containing a ``sea" of rural tracts, separated by, and surrounding, the urban centers. As such, both sets include some census tracts with no contiguous neighbors. In order to avoid inconsistencies in how spatial models treat units with no neighbors (and thus $0$ spatial error component) and units with neighbors, we choose a definition of neighboring that does not rely on spatial contiguity and takes into account the distance between tracts. Neighbors are defined by a threshold, chosen by determining the nearest neighbor distance for each tract using kNN with $k = 1$ and then taking the threshold to be the maximum nearest distance. \\

To generate the spatial weight matrix $\boldsymbol{W}$, we use the inverse distance for each neighbor. That is, we calculate the distance, $d$, from each census tract's population-weighted centroid to all other census tracts within the threshold. Each cell of the matrix $W_{ij}$ represents the weight for tract $j$ on tract $i$ and is equal to one over the distance between the two tracts for neighboring tracts and $0$ otherwise:
$$ w_{ij} = \begin{cases}
        1/d_{ij} & \text{if } d < \text{threshold}\\
        0 & \text{otherwise}
 \end{cases}
 $$ \\

This approach ensures that all tracts have at least one neighbor, but that distant tracts contribute relatively little importance. \\

\begin{figure}[H]
    \centering
    \vspace{-6em} 
    \hspace{-6em}
    \begin{minipage}{0.45\textwidth} 
        \centering
        \begin{subfigure}[b]{\textwidth}
            \centering
            \includegraphics[width=1.4\textwidth]{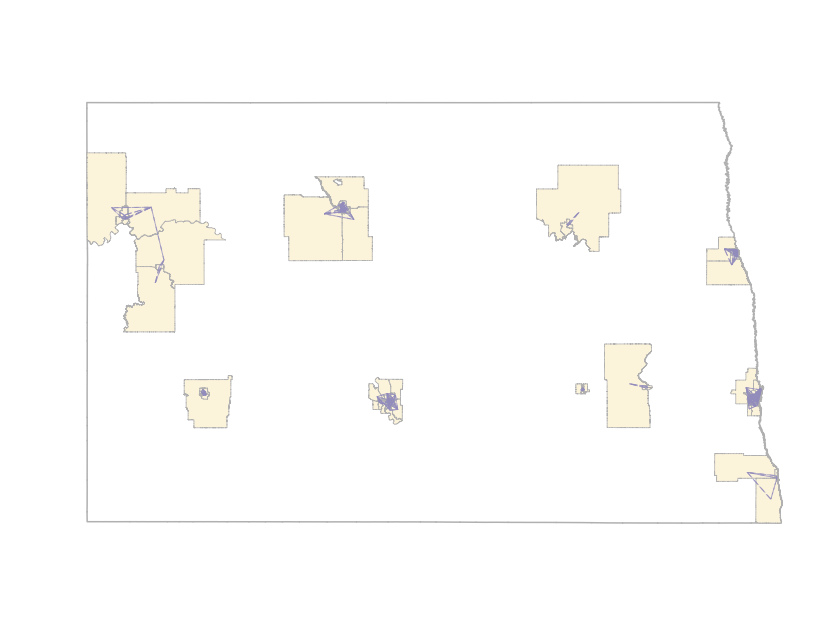}
            \caption{A}
        \end{subfigure}
        
        \begin{subfigure}[b]{\textwidth}
            \centering
            \includegraphics[width=1.4\textwidth]{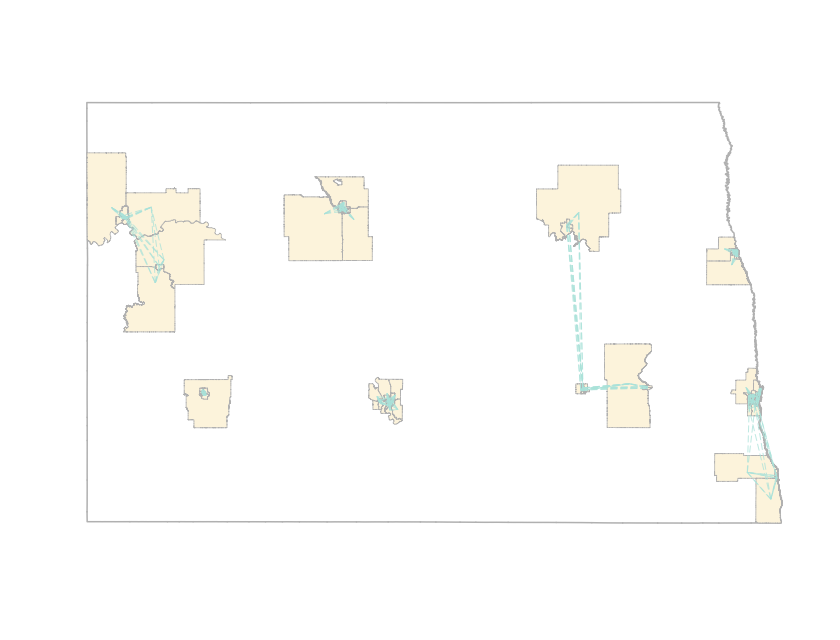}
            \caption{B}
        \end{subfigure}
        
        \begin{subfigure}[b]{\textwidth}
            \centering
            \includegraphics[width=1.4\textwidth]{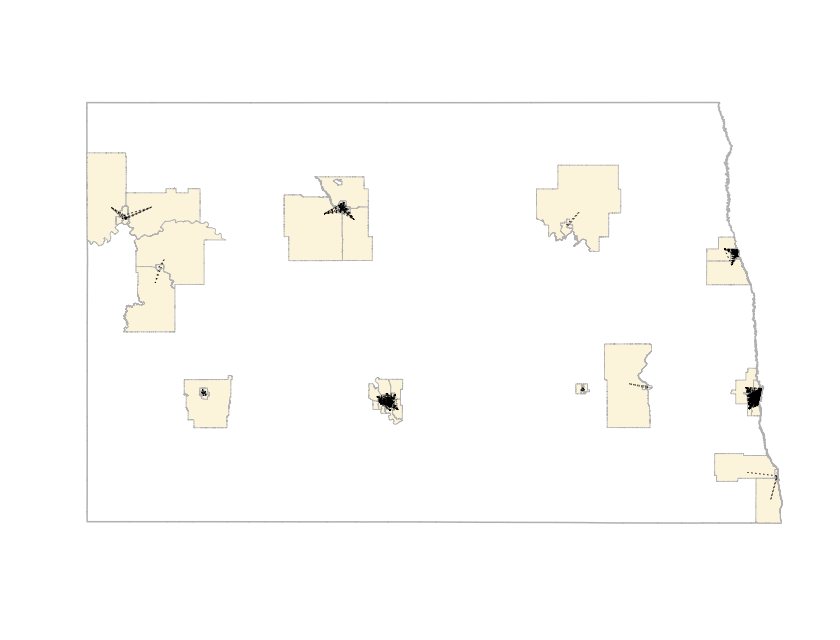}
            \caption{C}
        \end{subfigure}
    \end{minipage}
    \hfill
    \begin{minipage}{0.45\textwidth} 
        \centering
        \begin{subfigure}[b]{\textwidth}
            \centering
            \includegraphics[width=1.4\textwidth]{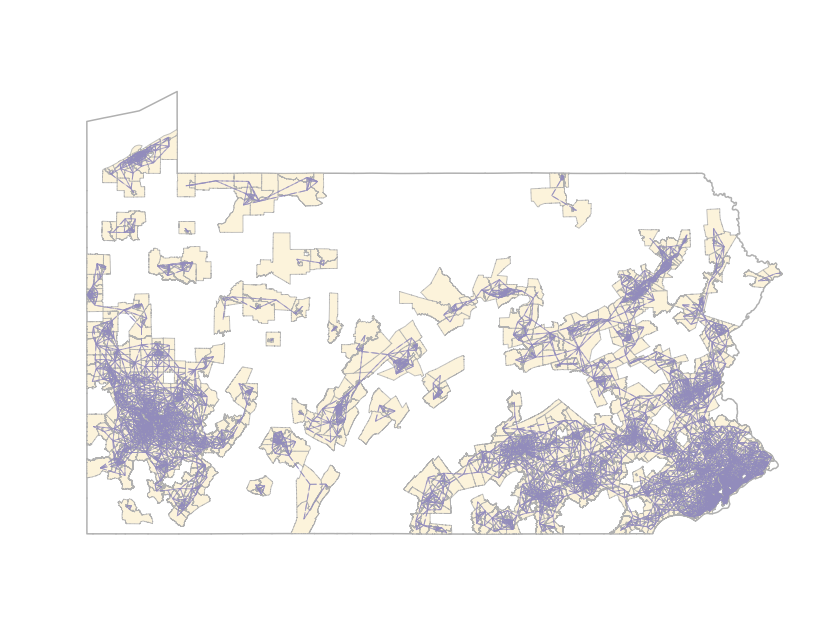}
            \caption{D}
        \end{subfigure}
        
        \begin{subfigure}[b]{\textwidth}
            \centering
            \includegraphics[width=1.4\textwidth]{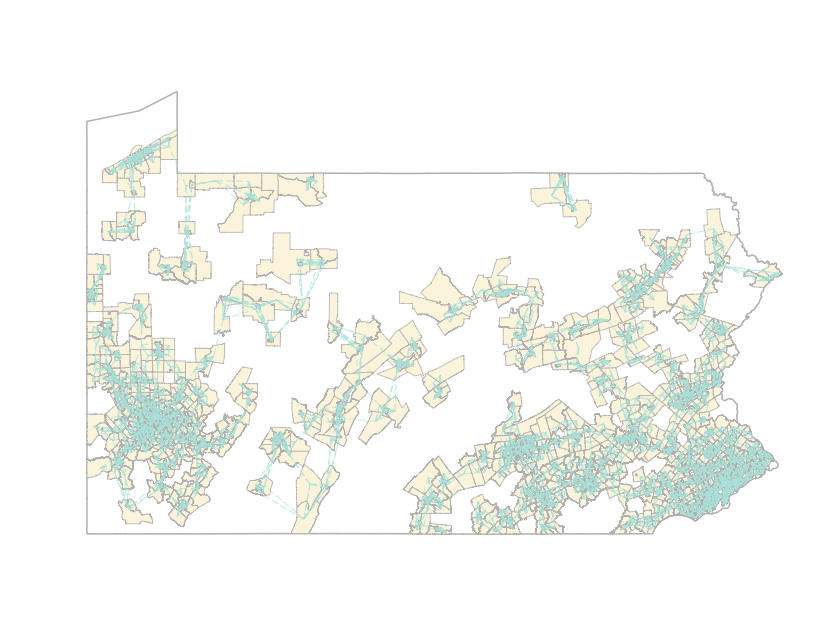}
            \caption{E}
        \end{subfigure}
        
        \begin{subfigure}[b]{\textwidth}
            \centering
            \includegraphics[width=1.4\textwidth]{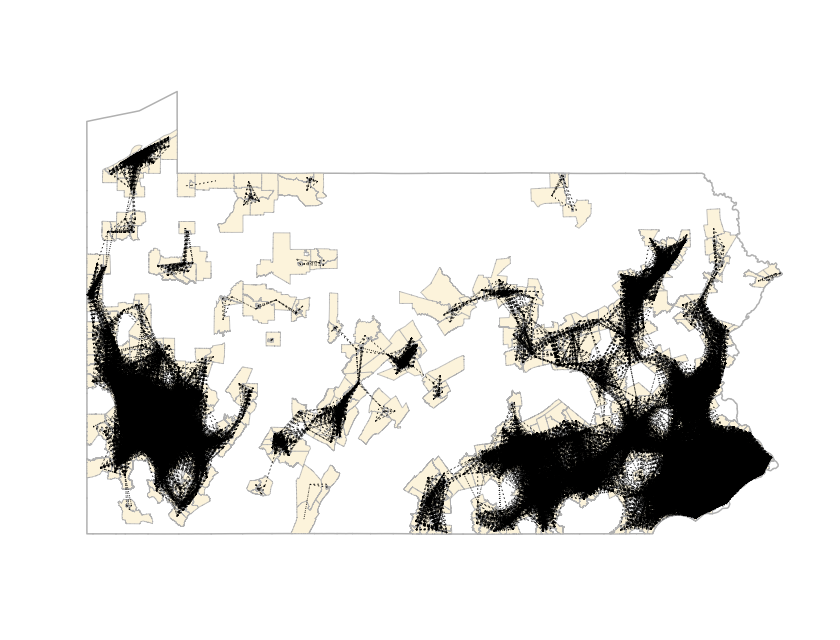}
            \caption{F}
        \end{subfigure}
    \end{minipage}
    \caption{Cream-colored urban census tracts in North Dakota (A-C) and Pennsylvania (D-F) are separated by areas of rural tracts (not displayed). Tracts identified as neighbors are linked by purple (spatially contiguous), blue (kNN, k = 5), and black (distance-threshold) lines. Notice that kNN defines some tracts from distinct urban clusters as neighbors in North Dakota.}
    \label{fig:neigh_urban}
\end{figure}

\begin{figure}[H] 
    \centering
    \vspace{-6em} 
    \hspace{-6em}
    \begin{minipage}{0.45\textwidth} 
        \centering
        \begin{subfigure}[b]{\textwidth}
            \centering
            \includegraphics[width=1.4\textwidth]{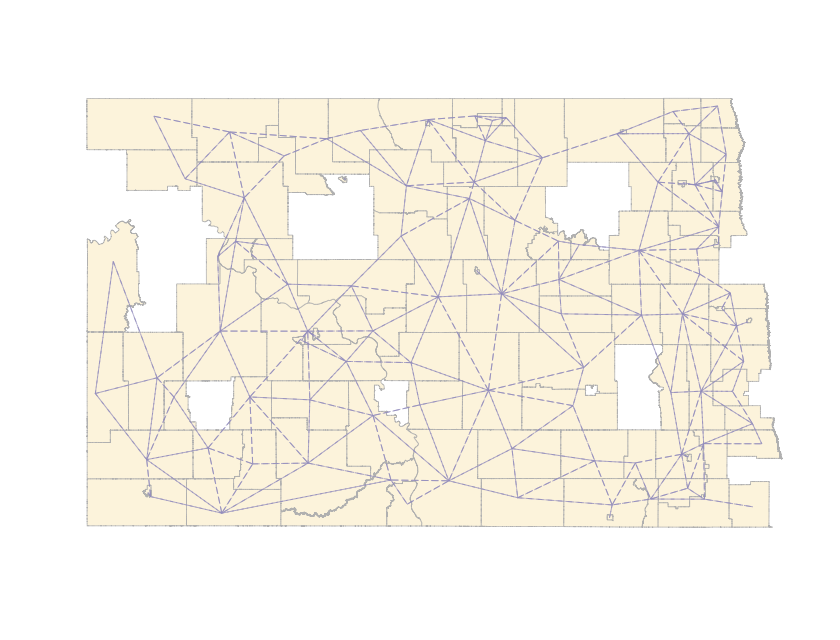}
            \caption{A}
        \end{subfigure}
        
        \begin{subfigure}[b]{\textwidth}
            \centering
            \includegraphics[width=1.4\textwidth]{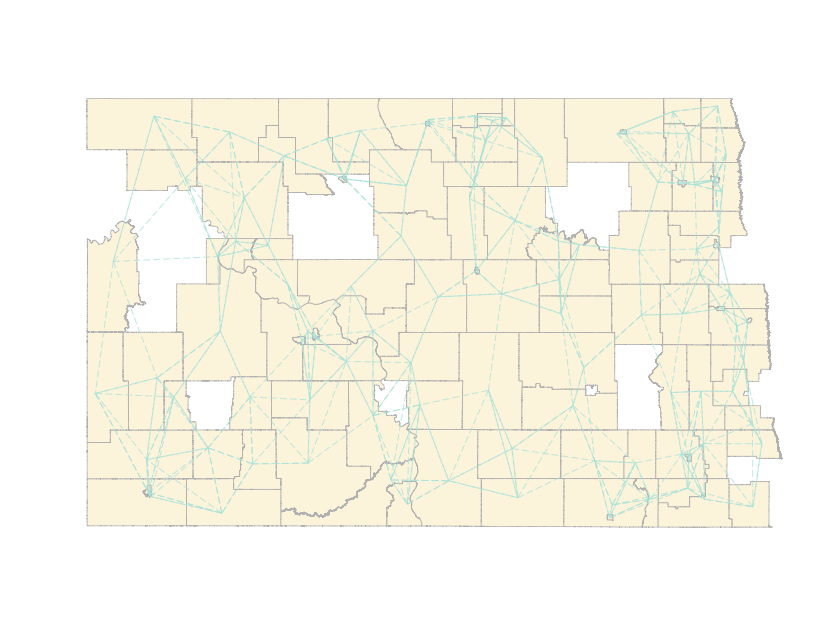}
            \caption{B}
        \end{subfigure}
        
        \begin{subfigure}[b]{\textwidth}
            \centering
            \includegraphics[width=1.4\textwidth]{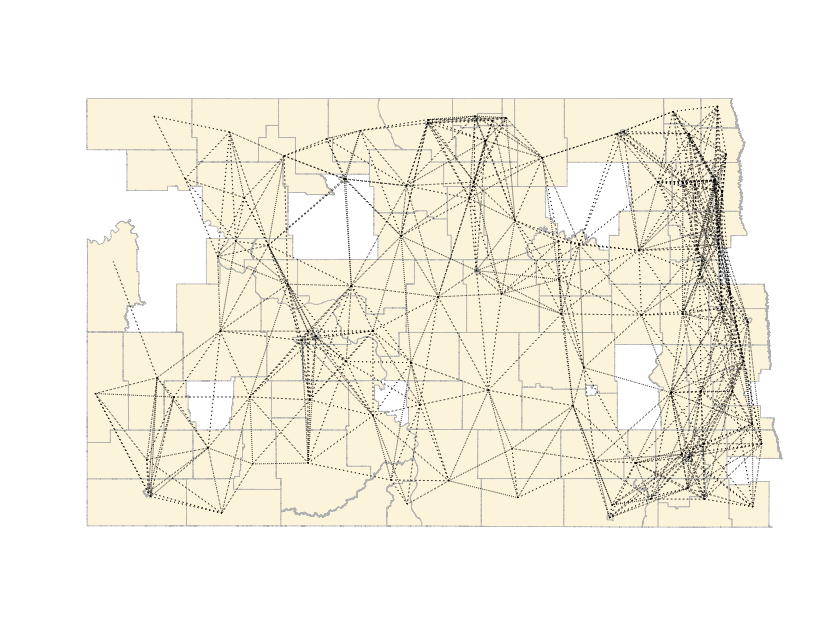}
            \caption{C}
        \end{subfigure}
    \end{minipage}
    \hfill
    \begin{minipage}{0.45\textwidth} 
        \centering
        \begin{subfigure}[b]{\textwidth}
            \centering
            \includegraphics[width=1.4\textwidth]{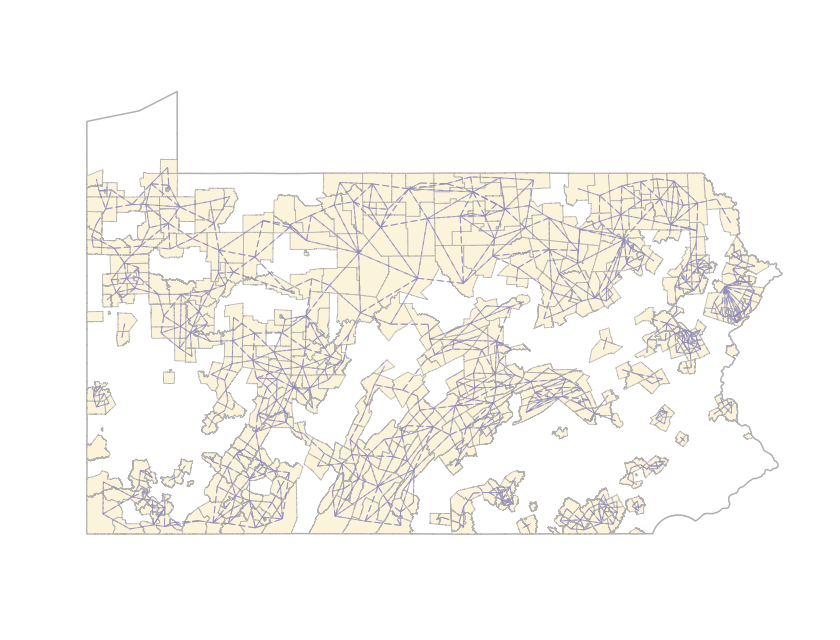}
            \caption{D}
        \end{subfigure}
        
        \begin{subfigure}[b]{\textwidth}
            \centering
            \includegraphics[width=1.4\textwidth]{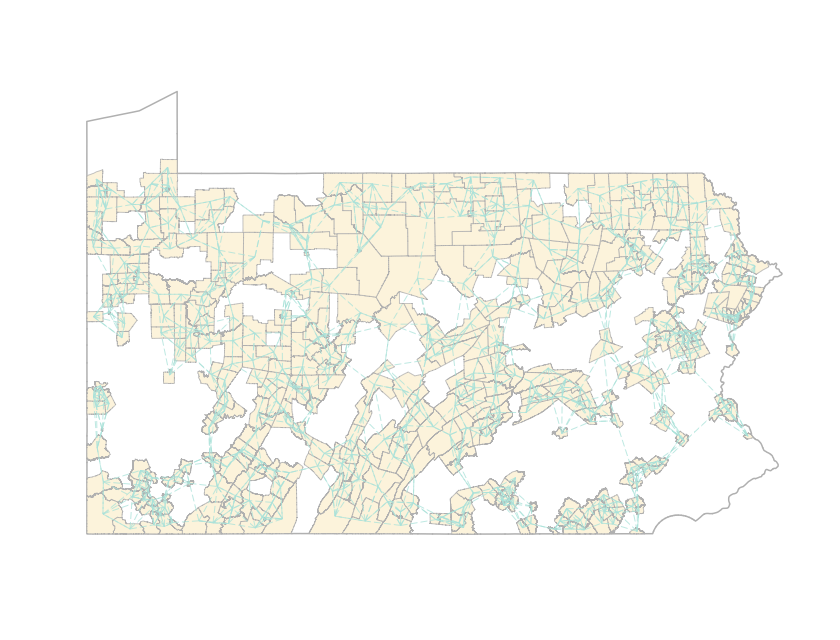}
            \caption{E}
        \end{subfigure}
        
        \begin{subfigure}[b]{\textwidth}
            \centering
            \includegraphics[width=1.4\textwidth]{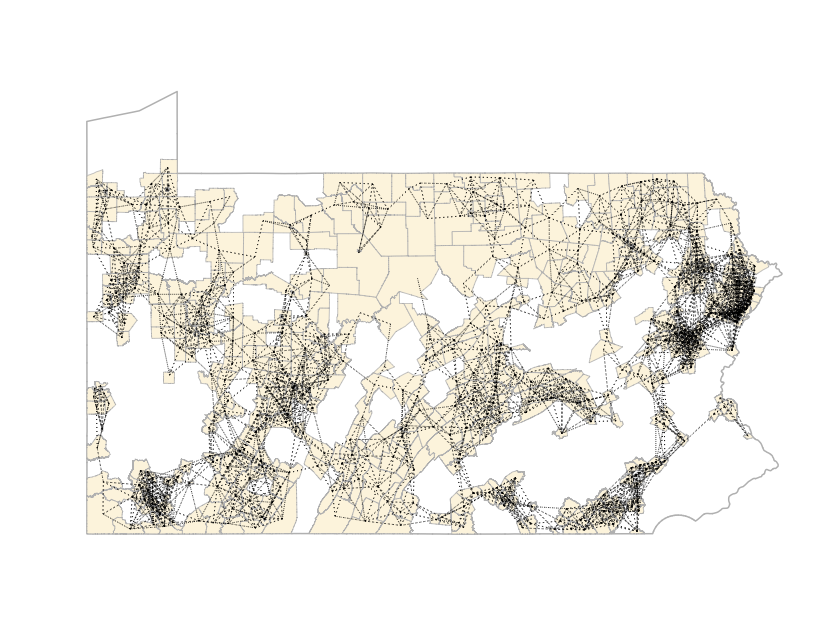}
            \caption{F}
        \end{subfigure}
    \end{minipage}
    \caption{Cream-colored rural census tracts in North Dakota (A-C) and Pennsylvania (D-F). Tracts identified as neighbors are linked by purple (spatially contiguous), blue (kNN, k = 5), and black (distance-threshold) lines. Notice that the spatially contiguous definition of neighboring tracts results in some tracts with zero neighbors in Pennsylvania.}
    \label{fig:neigh_rural}
\end{figure}

Other common methods for defining neighboring census tracts rely either on spatial contiguity or consistency in the size and separation of units \cite{getis_spatial_2010}\cite{bivand_modelling_2013}\cite{moraga_chapter_nodate}. Take for instance spatially contiguous neighbors with rook or queen-style contiguities. By this definition, census tracts are neighbors if they share a boundary edge or boundary point. Computational problems arise however when there are tracts which do not share a boundary with any other tracts (islands). Typically, these tracks are simply excluded from the analysis. Another common approach is kNN. However, the literature is limited on the best approach for determining the value for $k$.  Furthermore, the $k^{th}$ neighbor may be much further away for a geographically large, suburban census located on the outskirts of a densely populated area as compared to a smaller urban tract densely surrounded by other tracts. Neither a binary ($w_{ij} = 1$ for neighbors and $0$ otherwise) nor a row-standardized ($w_{ij} = 1/N$, for the total number of neighbors $N$) encoding of neighbors into the weight matrix takes into account the distances between tracts. As an example comparison between these three methods, figures \ref{fig:neigh_urban} and \ref{fig:neigh_rural} display neighboring tracts for urban and rural census tracts in North Dakota and Pennsylvania. In this example we consider $k = 5$ for the kNN. 

\newpage

\newpage 

\newpage

\subsubsection{Testing for spatial autocorrelation}\label{sec:moran}

We assess the spatial distribution of the multilevel model residuals by performing a global Moran’s I test \cite{getis_spatial_2010} (eq. \ref{eq:moranI}). The value of the Moran’s I test statistic varies between $-1$ (negative spatial autocorrelation) and $+1$ (positive spatial autocorrelation), with a value 0 indicating no autocorrelation. A statistically significant p-value indicates the presence of spatial autocorrelation in the residuals, suggesting that the multilevel model assumption that residuals are independent is not met, and justifying the use of a spatial model \cite{getis_spatial_2010}.

\begin{equation} \label{eq:moranI}
I = \frac{n}{\sum_{i = 1}^n \sum_{j = 1}^n W_{ij}} \frac{\boldsymbol{\epsilon}^T \boldsymbol{W} \boldsymbol{\epsilon}}{\boldsymbol{\epsilon}^T \boldsymbol{\epsilon}},
\end{equation}

Here $\boldsymbol{\epsilon}$ is the vector of regression residuals from the multilevel model, and $n$ is the total number of census tracts. $\boldsymbol{W}$ is the weight matrix defined in section \ref{sec:neighboring}.  

\subsubsection{Accounting for spatial clustering}\label{sec:INLA}

Including a random intercept at a larger spatial scale for units nested within larger geographic entities (census tracts within counties within states, for example) may or may not address the spatial dependencies in the data. We test for this by applying a global Moran's I statistic to the multilevel model residuals. If statistically significant, this indicates that a spatial model may be better suited for the data. \\

There are two main approaches to incorporating a spatial component into a model: including a spatial "lag" term or (and) a spatial "error" term (analogous to auto-regression or moving-averages in time series data). Although our data structure appears to be a substantive spatial process, for which de Graaff et al. recommend using a spatial lag \cite{de_graaff_general_2001}, Bivand \cite{bivand_modelling_2013} cautions against the interpretation of the fitted parameters $\beta$ as the expected impact of a one-unit change in the covariate on the outcome as the spatial lag model (unlike the spatial error model) results in feedback in the parameter $\rho_{Lag}$. Furthermore, the implication of the spatial lag model is that the outcome for one tract depends directly on the outcome in a neighboring tract \cite{bivand_modelling_2013} - while monitor proximity is certainly correlated, unlike with infectious disease dynamics it does not seem realistic that proximity for one tract would be a causal mechanism driving proximity for its neighbors. For this reason, and to ensure interpretability of the coefficient on the EJ attribute variable, we choose to model the structure using a spatial error term.  \\

To do so we take advantage of a Bayesian modelling technique that uses Integrated Nested Laplace Approximation (INLA) \cite{rue_approximate_2009} to incorporate a spatial model component alongside random intercepts in a multilevel model \cite{blangiardo_spatial_2013}. INLA is a computationally efficient alternative to Markov Chain Monte Carlo (MCMC) approximations with implementations available in R through the package \textit{R-INLA} \cite{rue_r-inla_2013}. INLA has been widely used in epidemiological studies to build multilevel spatial-temporal models with spatial error terms \cite{khana_bayesian_2018,satria_spatial_2021,teng_bayesian_2023}. \\

Following Tabb et al. \cite{tabb_spatially_2022}, we model the county and state-level random effects  as independent and identically distributed (i.i.d.) and the spatial error according to the Besag-York-Mollie (B-Y-M) model \cite{besag_bayesian_1991} (eq. \ref{eq:INLA}). Similar to the multilevel model, $Y_{ijk}$, the log-distance to the nearest monitor for census tract $i$ in county $j$ and state $k$, is assumed to be a linear function of the fixed intercept $\beta_0$, the EJ attribute, additional covariates $\vec{X}$, county ($\gamma_{jk}$) and state ($\alpha_k$)-level random effects with an additional error term composed of a spatially structured ($v_{ijk}$) and a spatially unstructured ($u_{ijk}$) component. \\

\begin{equation} \label{eq:INLA}
Y_{ijk} = \beta_0 + \beta_1 EJ + \vec{\beta_x} \vec{X} + \gamma_{jk} + \alpha_{k} + v_{ijk} + u_{ijk}
\end{equation}

The B-Y-M specification for the error term is a convolution of an i.i.d. Gaussian model and an intrinsic conditional autoregressive (iCAR) \cite{gomez-rubio_bayesian_2020}, where the prior for unstructured error follows the usual specification of being normally distributed with a mean of $0$ and constant variance $\sigma_u^2$ and the prior for the structured error, $v_{ijk}$, follows a normal distribution with mean $m_i$ and variance $s_i^2$ \cite{blangiardo_spatial_2013}\cite{congdon_applied_2014} (eq. \ref{eq:BYM}). Where $m_i$ for census tract $i$ is a function of the spatial error, $v_{ljk}$, of its neighboring census tracts $l$, and the neighborhood weight matrix $W_{il}$, defined previously. Conceptually, the interpretation of this model is that the estimates at any given location are conditional on the level of neighboring values. \\

\begin{equation} \label{eq:BYM}
m_i = \frac{\sum_{l=1}^n w_{il} v_{ljk}}{\sum_{l=1}^n w_{il}} \text{ ; }
s_i^2 = \frac{\sigma_v^2}{\sum_{l=1}^n w_{il}}
\end{equation}

Random intercepts $\gamma_{jk}$ and $\alpha_k$ are modelled as i.i.d. in \textit{INLA}\cite{gomez-rubio_bayesian_2020}. We use the default minimally informative prior of $logGamma(1, 0.0005)$ for the hyperparameter precision for all parameters (note that precision is the inverse of variance, $\tau = 1/\sigma^2$) \cite{gomez-rubio_bayesian_2020}. The neighborhood matrix $W$ was constructed using the R package \textit{spdep} \cite{bivand_modelling_2013}. To reduce computational burden (a $70{,}000$ x $70{,}000$ weight matrix representing the relationship between every urban census tract in the US uses approximately $80$ GB of memory), we use the \textit{CsparseMatrix} representation from the R package \textit{Matrix} \cite{bates_matrix_2024} for $W$. We retrieve the posterior mean and 95\% credible intervals for fixed and random effects and compared these estimates from the Bayesian INLA model with and without a spatial error term, and to the frequentist multilevel models for each EJ attribute-model. All statistical analyses were conducted in R version 4.4.0 \cite{r_core_team_r_2014}.

\subsection{Results}

\subsubsection{Moran's I test for spatial autocorrelation in model residuals}

\begin{table}[ht]
\centering
    \caption{Moran's I test on US-wide multilevel model residuals.}
    \label{tb:moran}
\begin{NiceTabular}{ccc}[vlines] \hline
    
        \Block{2-1}{EJ attribute model} & \Block{1-2}{\textbf{Moran's I test results}}\\ 
         & Rural & Urban \\ \hline
        \Block{4-1}{\%AIAN} & Moran I statistic = 4.817738e-02 & Moran I statistic = 1.664633e-01 \\ 
                            & Expectation = -8.442381e-05 & Expectation = -1.418802e-05\\ 
                            & Variance = 2.363839e-06 & Variance = 3.123668e-07\\ 
                            & p-value $<$ 2.2e-16 & p-value $<$ 2.2e-16 \\ \hline
        \Block{4-1}{\%Asian} & Moran I statistic = 4.818524e-02 & Moran I statistic = 1.664880e-01 \\ 
                            & Expectation =-8.442381e-05 & Expectation = -1.418802e-05\\ 
                            & Variance = 2.363828e-06 & Variance = 3.123668e-07\\ 
                          & p-value $<$ 2.2e-16 & p-value $<$ 2.2e-16\\ \hline
        \Block{4-1}{\%Black} & Moran I statistic = 4.794237e-02 & Moran I statistic = 1.661750e-01\\ 
                        & Expectation =-8.442381e-05 & Expectation = -1.418802e-05\\ 
                        & Variance = 2.363830e-06 & Variance = 3.123668e-07\\ 
                        & p-value $<$ 2.2e-16 & p-value $<$ 2.2e-16 \\ \hline
        \Block{4-1}{\%Hispanic} & Moran I statistic = 4.816147e-02 & Moran I statistic = 1.658527e-01 \\ 
                        & Expectation = -8.442381e-05 & Expectation =-1.418802e-05 \\ 
                        & Variance = 2.363831e-06 & Variance = 3.123668e-07 \\ 
                        & p-value $<$ 2.2e-16 & p-value $<$ 2.2e-16 \\ \hline
        \Block{4-1}{\%Poverty} & Moran I statistic = 4.816533e-02 & Moran I statistic = 1.664752e-01 \\ 
                        & Expectation = -8.442381e-05 & Expectation = -1.418802e-05\\ 
                        & Variance = 2.363831e-06 & Variance = 3.123668e-07\\ 
                        & p-value $<$ 2.2e-16 & p-value $<$ 2.2e-16 \\ \hline
        \Block{4-1}{\%White} & Moran I statistic = 4.816533e-02  & Moran I statistic = 1.664752e-01\\ 
                        & Expectation =-8.442381e-05 & Expectation = -1.418802e-05\\ 
                        & Variance = 2.363831e-06 & Variance = 3.123668e-07\\ 
                        & p-value $<$ 2.2e-16 & p-value $<$ 2.2e-16 \\ \hline

\end{NiceTabular}
\end{table}

Table \ref{tb:moran} displays the Moran's I test statistics for spatial autocorrelation of model residuals for the US-wide multilevel models. For both urban and rural tracts, we detect a statistically significant spatial autocorrelation ($p < 0.05$) among model residuals for all EJ attributes. 

\newpage

\subsubsection{Comparison between multilevel and spatial error models}

\begin{figure}[h]
\includegraphics[width=1.2\linewidth]{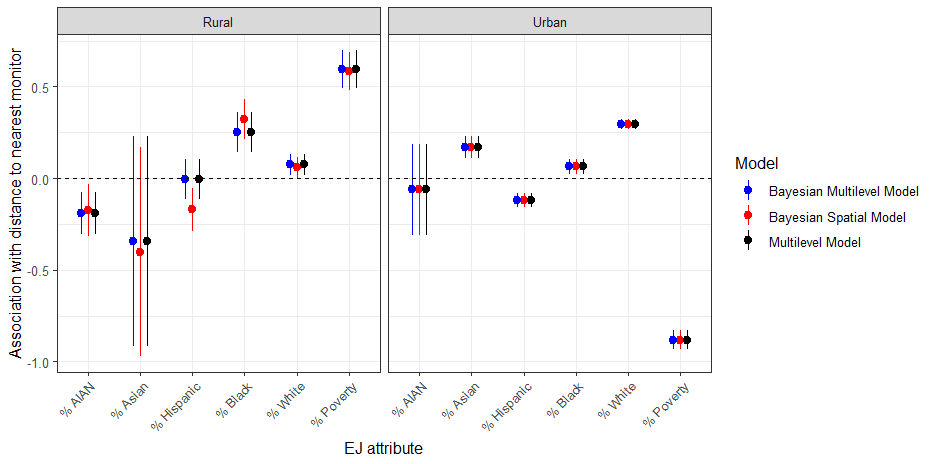}
\caption{The association between the EJ attribute and the distance to the nearest PM monitor for rural and urban tracts when modelled using a Bayesian multilevel (blue) or Bayesian multilevel with spatial error term (red) \textit{INLA} approach, or the frequentist multilevel model (black). Circles indicate point estimates and vertical bars the 95\% credible intervals (Bayesian models) or 95\% confidence intervals (frequentist model).}
\label{fig:INLA_comparison}
\end{figure}

As with the non-spatial model, we construct a separate Bayesian INLA model for each EJ attribute under consideration. We compare the estimate of the coefficient on the EJ attribute for the Bayesian models with and without a spatial error term, and to the estimates from the non-spatial multilevel model (figure \ref{fig:INLA_comparison}). Point estimates and 95\% confidence intervals are nearly identical between all three models for all EJ attributes.

\newpage


\newpage

\renewcommand{\thesection}{Appendix \Alph{section}}
\section{Supporting Materials}




\renewcommand{\thesection}{\Alph{section}}

\subsection{List of multilevel models considered}

All models were run separately for urban and rural census tracts. \\

\subsubsection{Main analysis}\label{sec:main_models}
EJ attribute: SES operationalized as the proportion of population living below the poverty line
\begin{equation} \label{eq. 1}
    \begin{split}
    Y_{ijk} &= \beta_{0} + \beta_{1} \text{Prop. Poverty} + \\
    & \beta_2 \text{Prop non-White} + \beta_3 \text{Population density} + \beta_4 \text{PM2.5 Z-score} + \\
    & \gamma_{jk} + \alpha_{k} + \epsilon_{ijk}
\end{split}
\end{equation}

EJ attribute: proportion of census tract identifying as American Indian or Alaskan Native (AIAN)
\begin{equation} \label{eq. 2}
    \begin{split}
    Y_{ijk} &= \beta_{0} + \beta_{1}\text{Prop. AIAN}  + \\
    & \beta_{2}\text{Prop. White} + \beta_3\text{Prop. Poverty}  + \beta_4 \text{Population density} + \\
    & \beta_5 \text{PM2.5 Z-score} + \gamma_{jk} + \alpha_{k} + \epsilon_{ijk}
\end{split}
\end{equation}

EJ attribute: proportion of census tract identifying as Asian
\begin{equation} \label{eq. 3}
    \begin{split}
    Y_{ijk} &= \beta_{0} + \beta_{1}\text{Prop. Asian}  + \\
    & \beta_{2}\text{Prop. White} + \beta_3\text{Prop. Poverty}  + \beta_4 \text{Population density} + \\
    &\beta_5 \text{PM2.5 Z-score} +  \gamma_{jk} + \alpha_{k} + \epsilon_{ijk}
\end{split}
\end{equation}

EJ attribute: proportion of census tract identifying as Black
\begin{equation} \label{eq. 4}
    \begin{split}
    Y_{ijk} &= \beta_{0} + \beta_{1}\text{Prop. Black}  + \\
    & \beta_{2}\text{Prop. White} + \beta_3\text{Prop. Poverty}  + \beta_4 \text{Population density} + \\
    & \beta_5 \text{PM2.5 Z-score} +  \gamma_{jk} + \alpha_{k} + \epsilon_{ijk}
\end{split}
\end{equation}

EJ attribute: proportion of census tract identifying as Hispanic
\begin{equation} \label{eq. 5}
    \begin{split}
    Y_{ijk} &= \beta_{0} + \beta_{1}\text{Prop. Hispanic}  + \\
    & \beta_{2}\text{Prop. White} + \beta_3\text{Prop. Poverty}  + \beta_4 \text{Population density} + \\
    & \beta_5 \text{PM2.5 Z-score} +  \gamma_{jk} + \alpha_{k} + \epsilon_{ijk}
\end{split}
\end{equation}

EJ attribute: proportion of census tract identifying as White
\begin{equation} \label{eq. 6}
    \begin{split}
    Y_{ijk} &= \beta_{0} + \beta_{1}\text{Prop. White}  + \\
    & \beta_2\text{Prop. Poverty}  + \beta_3 \text{Population density} + \beta_4 \text{PM2.5 Z-score} + \\
    & \gamma_{jk} + \alpha_{k} + \epsilon_{ijk}
\end{split}
\end{equation}

\subsubsection{Sensitivity analysis: SES operationalized as median household income}

EJ attribute: median household income
\begin{equation} \label{eq. 7}
    \begin{split}
    Y_{ijk} &= \beta_{0} + \beta_{1} \text{Household Income (in 100k)} + \\
    & \beta_2 \text{Prop non-White} + \beta_3 \text{Population density} + \beta_4 \text{PM2.5 Z-score} + \\
    & \gamma_{jk} + \alpha_{k} + \epsilon_{ijk}
\end{split}
\end{equation}

EJ attribute: proportion of census tract identifying as American Indian or Alaskan Native (AIAN)
\begin{equation} \label{eq. 8}
    \begin{split}
    Y_{ijk} &= \beta_{0} + \beta_{1}\text{Prop. AIAN}  + \\
    & \beta_{2}\text{Prop. White} + \beta_3\text{Household Income (in 100k)}  + \beta_4 \text{Population density} + \\
    & \beta_5 \text{PM2.5 Z-score} +  \gamma_{jk} + \alpha_{k} + \epsilon_{ijk}
\end{split}
\end{equation}

EJ attribute: proportion of census tract identifying as Asian
\begin{equation} \label{eq. 9}
    \begin{split}
    Y_{ijk} &= \beta_{0} + \beta_{1}\text{Prop. Asian}  + \\
    & \beta_{2}\text{Prop. White} + \beta_3\text{Household Income (in 100k)}  + \beta_4 \text{Population density} + \\
    & \beta_5 \text{PM2.5 Z-score} + \gamma_{jk} + \alpha_{k} + \epsilon_{ijk}
\end{split}
\end{equation}

EJ attribute: proportion of census tract identifying as Black
\begin{equation} \label{eq. 10}
    \begin{split}
    Y_{ijk} &= \beta_{0} + \beta_{1}\text{Prop. Black}  + \\
    & \beta_{2}\text{Prop. White} + \beta_3\text{Household Income (in 100k)}  + \beta_4 \text{Population density} + \\
    & \beta_5 \text{PM2.5 Z-score} + \gamma_{jk} + \alpha_{k} + \epsilon_{ijk}
\end{split}
\end{equation}

EJ attribute: proportion of census tract identifying as Hispanic
\begin{equation} \label{eq. 11}
    \begin{split}
    Y_{ijk} &= \beta_{0} + \beta_{1}\text{Prop. Hispanic}  + \\
    & \beta_{2}\text{Prop. White} + \beta_3\text{Household Income (in 100k)}  + \beta_4 \text{Population density} + \\
    & \beta_5 \text{PM2.5 Z-score} + \gamma_{jk} + \alpha_{k} + \epsilon_{ijk}
\end{split}
\end{equation}

EJ attribute: proportion of census tract identifying as White
\begin{equation} \label{eq. 12}
    \begin{split}
    Y_{ijk} &= \beta_{0} + \beta_{1}\text{Prop. White}  + \\
    & \beta_2\text{Household Income (in 100k)}  + \beta_3 \text{Population density} + \beta_4 \text{PM2.5 Z-score} + \\
    & \gamma_{jk} + \alpha_{k} + \epsilon_{ijk}
\end{split}
\end{equation}

\newpage

\subsubsection{Multilevel model coefficient estimates}\label{sec:sensitivity_models}

\begin{landscape}
    \begin{table}[!htbp] \centering 
  \caption{Rural} 
  \label{tb:mlm_rural} 
\begin{tabular}{@{\extracolsep{5pt}}lD{.}{.}{-3} D{.}{.}{-3} D{.}{.}{-3} D{.}{.}{-3} D{.}{.}{-3} D{.}{.}{-3} } 
\\[-1.8ex]\hline 
\hline \\[-1.8ex] 
 & \multicolumn{6}{c}{\textit{EJ Attribute:}} \\ 
\cline{2-7} 
\\[-1.8ex] & \multicolumn{1}{c}{\%AIAN} & \multicolumn{1}{c}{\%Asian} & \multicolumn{1}{c}{\%Black} & \multicolumn{1}{c}{\%Hispanic} & \multicolumn{1}{c}{\%Below Poverty} & \multicolumn{1}{c}{\%White}\\ 
\hline \\[-1.8ex] 
 Prop. Poverty & 0.611^{***} & 0.592^{***} & 0.594^{***} & 0.595^{***} & 0.596^{***} & 0.596^{***} \\ 
  & (0.054) & (0.054) & (0.053) & (0.054) & (0.053) & (0.053) \\ 
  Prop. AIAN & -0.191^{***} &  &  &  &  &  \\ 
  & (0.059) &  &  &  &  &  \\ 
  Prop. Asian &  & -0.343 &  &  &  &  \\ 
  &  & (0.292) &  &  &  &  \\ 
  Prop. Black &  &  & 0.250^{***} &  &  &  \\ 
  &  &  & (0.055) &  &  &  \\ 
  Prop. Hispanic &  &  &  & -0.006 &  &  \\ 
  &  &  &  & (0.055) &  &  \\ 
  Prop. White & 0.031 & 0.073^{**} & 0.186^{***} & 0.074^{**} &  & 0.075^{**} \\ 
  & (0.033) & (0.030) & (0.038) & (0.034) &  & (0.030) \\ 
  Prop. nonWhite &  &  &  &  & -0.075^{**} &  \\ 
  &  &  &  &  & (0.030) &  \\ 
  Pop. Density & -3.077^{***} & -2.937^{***} & -3.019^{***} & -2.964^{***} & -2.965^{***} & -2.965^{***} \\ 
  & (0.481) & (0.481) & (0.480) & (0.480) & (0.480) & (0.480) \\ 
  PM2.5 Z Score & -0.265^{***} & -0.262^{***} & -0.257^{***} & -0.262^{***} & -0.262^{***} & -0.262^{***} \\ 
  & (0.013) & (0.013) & (0.013) & (0.013) & (0.013) & (0.013) \\ 
  Intercept & 10.339^{***} & 10.306^{***} & 10.201^{***} & 10.304^{***} & 10.377^{***} & 10.301^{***} \\ 
  & (0.061) & (0.060) & (0.064) & (0.063) & (0.055) & (0.060) \\ 
 \hline \\[-1.8ex] 
Observations & \multicolumn{1}{c}{11,846} & \multicolumn{1}{c}{11,846} & \multicolumn{1}{c}{11,846} & \multicolumn{1}{c}{11,846} & \multicolumn{1}{c}{11,846} & \multicolumn{1}{c}{11,846} \\  
\hline 
\hline \\[-1.8ex] 
\textit{Note:}  & \multicolumn{6}{r}{$^{*}$p$<$0.1; $^{**}$p$<$0.05; $^{***}$p$<$0.01} \\ 
\end{tabular} 
\end{table}
\end{landscape}

\begin{landscape}
\begin{table}[!htbp] \centering 
  \caption{Urban} 
  \label{tb:mlm_urban} 
\begin{tabular}{@{\extracolsep{5pt}}lD{.}{.}{-3} D{.}{.}{-3} D{.}{.}{-3} D{.}{.}{-3} D{.}{.}{-3} D{.}{.}{-3} } 
\\[-1.8ex]\hline 
\hline \\[-1.8ex] 
 & \multicolumn{6}{c}{\textit{EJ Attribute:}} \\ 
\cline{2-7} 
\\[-1.8ex] & \multicolumn{1}{c}{\%AIAN} & \multicolumn{1}{c}{\%Asian} & \multicolumn{1}{c}{\%Black} & \multicolumn{1}{c}{\%Hispanic} & \multicolumn{1}{c}{\% Below Poverty} & \multicolumn{1}{c}{\%White}\\ 
\hline \\[-1.8ex] 
 Prop. Poverty & -0.882^{***} & -0.868^{***} & -0.892^{***} & -0.891^{***} & -0.883^{***} & -0.883^{***} \\ 
  & (0.027) & (0.027) & (0.027) & (0.027) & (0.027) & (0.027) \\ 
  Prop. AIAN & -0.061 &  &  &  &  &  \\ 
  & (0.126) &  &  &  &  &  \\ 
  Prop. Asian &  & 0.171^{***} &  &  &  &  \\ 
  &  & (0.031) &  &  &  &  \\ 
  Prop. Black &  &  & 0.064^{***} &  &  &  \\ 
  &  &  & (0.020) &  &  &  \\ 
  Prop. Hispanic &  &  &  & -0.121^{***} &  &  \\ 
  &  &  &  & (0.019) &  &  \\ 
  Prop. White & 0.295^{***} & 0.304^{***} & 0.331^{***} & 0.249^{***} &  & 0.296^{***} \\ 
  & (0.013) & (0.013) & (0.017) & (0.015) &  & (0.013) \\ 
  Prop. nonWhite &  &  &  &  & -0.296^{***} &  \\ 
  &  &  &  &  & (0.013) &  \\ 
  Pop. Density & -0.197^{***} & -0.198^{***} & -0.194^{***} & -0.191^{***} & -0.197^{***} & -0.197^{***} \\ 
  & (0.008) & (0.008) & (0.008) & (0.008) & (0.008) & (0.008) \\ 
  PM2.5 Z Score & -0.320^{***} & -0.319^{***} & -0.317^{***} & -0.315^{***} & -0.319^{***} & -0.319^{***} \\ 
  & (0.006) & (0.006) & (0.006) & (0.006) & (0.006) & (0.006) \\ 
  Intercept & 9.811^{***} & 9.798^{***} & 9.780^{***} & 9.861^{***} & 10.105^{***} & 9.810^{***} \\ 
  & (0.078) & (0.078) & (0.079) & (0.079) & (0.078) & (0.078) \\ 
 \hline \\[-1.8ex] 
Observations & \multicolumn{1}{c}{70,483} & \multicolumn{1}{c}{70,483} & \multicolumn{1}{c}{70,483} & \multicolumn{1}{c}{70,483} & \multicolumn{1}{c}{70,483} & \multicolumn{1}{c}{70,483} \\ 
\hline 
\hline \\[-1.8ex] 
\textit{Note:}  & \multicolumn{6}{r}{$^{*}$p$<$0.1; $^{**}$p$<$0.05; $^{***}$p$<$0.01} \\ 
\end{tabular} 
\end{table} 
\end{landscape}

\newpage 

\subsection{Multilevel model coefficient estimates - sensitivity analysis }

\begin{landscape}
\begin{table}[!htbp] \centering 
  \caption{Sensitivity Analysis Rural} 
  \label{tb:mlmS_rural} 
\begin{tabular}{@{\extracolsep{5pt}}lD{.}{.}{-3} D{.}{.}{-3} D{.}{.}{-3} D{.}{.}{-3} D{.}{.}{-3} D{.}{.}{-3} } 
\\[-1.8ex]\hline 
\hline \\[-1.8ex] 
 & \multicolumn{6}{c}{\textit{EJ Attribute:}} \\ 
\cline{2-7} 
\\[-1.8ex] & \multicolumn{1}{c}{\%AIAN} & \multicolumn{1}{c}{\%Asian} & \multicolumn{1}{c}{\%Black} & \multicolumn{1}{c}{\%Hispanic} & \multicolumn{1}{c}{HH Income} & \multicolumn{1}{c}{\%White}\\ 
\hline \\[-1.8ex] 
 Median HH Income (100k) & -0.571^{***} & -0.571^{***} & -0.567^{***} & -0.570^{***} & -0.570^{***} & -0.570^{***} \\ 
  & (0.027) & (0.027) & (0.027) & (0.027) & (0.027) & (0.027) \\ 
  Prop. AIAN & -0.149^{**} &  &  &  &  &  \\ 
  & (0.058) &  &  &  &  &  \\ 
  Prop. Asian &  & 0.138 &  &  &  &  \\ 
  &  & (0.290) &  &  &  &  \\ 
  Prop. Black &  &  & 0.228^{***} &  &  &  \\ 
  &  &  & (0.054) &  &  &  \\ 
  Prop. Hispanic &  &  &  & -0.044 &  &  \\ 
  &  &  &  & (0.054) &  &  \\ 
  Prop. White & 0.100^{***} & 0.138^{***} & 0.237^{***} & 0.124^{***} &  & 0.137^{***} \\ 
  & (0.032) & (0.029) & (0.038) & (0.033) &  & (0.029) \\ 
  Prop. nonWhite &  &  &  &  & -0.137^{***} &  \\ 
  &  &  &  &  & (0.029) &  \\ 
  Pop. Density & -3.329^{***} & -3.260^{***} & -3.293^{***} & -3.239^{***} & -3.247^{***} & -3.247^{***} \\ 
  & (0.477) & (0.477) & (0.476) & (0.476) & (0.476) & (0.476) \\ 
  PM2.5 Z Score & -0.255^{***} & -0.253^{***} & -0.249^{***} & -0.251^{***} & -0.253^{***} & -0.253^{***} \\ 
  & (0.013) & (0.013) & (0.013) & (0.013) & (0.013) & (0.013) \\ 
  Intercept & 10.716^{***} & 10.683^{***} & 10.591^{***} & 10.698^{***} & 10.820^{***} & 10.684^{***} \\ 
  & (0.056) & (0.054) & (0.059) & (0.057) & (0.052) & (0.054) \\ 
 \hline \\[-1.8ex] 
Observations & \multicolumn{1}{c}{11,846} & \multicolumn{1}{c}{11,846} & \multicolumn{1}{c}{11,846} & \multicolumn{1}{c}{11,846} & \multicolumn{1}{c}{11,846} & \multicolumn{1}{c}{11,846} \\ 
\hline 
\hline \\[-1.8ex] 
\textit{Note:}  & \multicolumn{6}{r}{$^{*}$p$<$0.1; $^{**}$p$<$0.05; $^{***}$p$<$0.01} \\ 
\end{tabular} 
\end{table} 
\end{landscape}

\begin{landscape}
    \begin{table}[!htbp] \centering 
  \caption{Sensitivity Analysis Urban} 
  \label{tb:mlmS_urban} 
\begin{tabular}{@{\extracolsep{5pt}}lD{.}{.}{-3} D{.}{.}{-3} D{.}{.}{-3} D{.}{.}{-3} D{.}{.}{-3} D{.}{.}{-3} } 
\\[-1.8ex]\hline 
\hline \\[-1.8ex] 
 & \multicolumn{6}{c}{\textit{EJ Attribute:}} \\ 
\cline{2-7} 
\\[-1.8ex] & \multicolumn{1}{c}{\%AIAN} & \multicolumn{1}{c}{\%Asian} & \multicolumn{1}{c}{\%Black} & \multicolumn{1}{c}{\%Hispanic} & \multicolumn{1}{c}{HH Income} & \multicolumn{1}{c}{\%White}\\ 
\hline \\[-1.8ex] 
 Median HH Income (100k) & 0.195^{***} & 0.186^{***} & 0.195^{***} & 0.193^{***} & 0.195^{***} & 0.195^{***} \\ 
  & (0.009) & (0.010) & (0.009) & (0.009) & (0.009) & (0.009) \\ 
  Prop. AIAN & -0.098 &  &  &  &  &  \\ 
  & (0.127) &  &  &  &  &  \\ 
  Prop. Asian &  & 0.139^{***} &  &  &  &  \\ 
  &  & (0.032) &  &  &  &  \\ 
  Prop. Black &  &  & 0.007 &  &  &  \\ 
  &  &  & (0.020) &  &  &  \\ 
  Prop. Hispanic &  &  &  & -0.048^{**} &  &  \\ 
  &  &  &  & (0.019) &  &  \\ 
  Prop. White & 0.334^{***} & 0.346^{***} & 0.339^{***} & 0.319^{***} &  & 0.335^{***} \\ 
  & (0.014) & (0.014) & (0.018) & (0.015) &  & (0.013) \\ 
  Prop. nonWhite &  &  &  &  & -0.335^{***} &  \\ 
  &  &  &  &  & (0.013) &  \\ 
  Pop. Density & -0.206^{***} & -0.207^{***} & -0.205^{***} & -0.204^{***} & -0.206^{***} & -0.206^{***} \\ 
  & (0.008) & (0.008) & (0.008) & (0.008) & (0.008) & (0.008) \\ 
  PM2.5 Z Score & -0.315^{***} & -0.315^{***} & -0.314^{***} & -0.313^{***} & -0.315^{***} & -0.315^{***} \\ 
  & (0.006) & (0.006) & (0.006) & (0.006) & (0.006) & (0.006) \\ 
  Intercept & 9.541^{***} & 9.533^{***} & 9.535^{***} & 9.558^{***} & 9.874^{***} & 9.539^{***} \\ 
  & (0.078) & (0.078) & (0.078) & (0.078) & (0.078) & (0.078) \\ 
 \hline \\[-1.8ex] 
Observations & \multicolumn{1}{c}{70,483} & \multicolumn{1}{c}{70,483} & \multicolumn{1}{c}{70,483} & \multicolumn{1}{c}{70,483} & \multicolumn{1}{c}{70,483} & \multicolumn{1}{c}{70,483} \\ 
\hline 
\hline \\[-1.8ex] 
\textit{Note:}  & \multicolumn{6}{r}{$^{*}$p$<$0.1; $^{**}$p$<$0.05; $^{***}$p$<$0.01} \\ 
\end{tabular} 
\end{table} 
\end{landscape}

\subsection{EJ attribute association with monitor proximity by \%poverty and median household income}

\begin{figure}[h]
\includegraphics[width=1.2\linewidth]{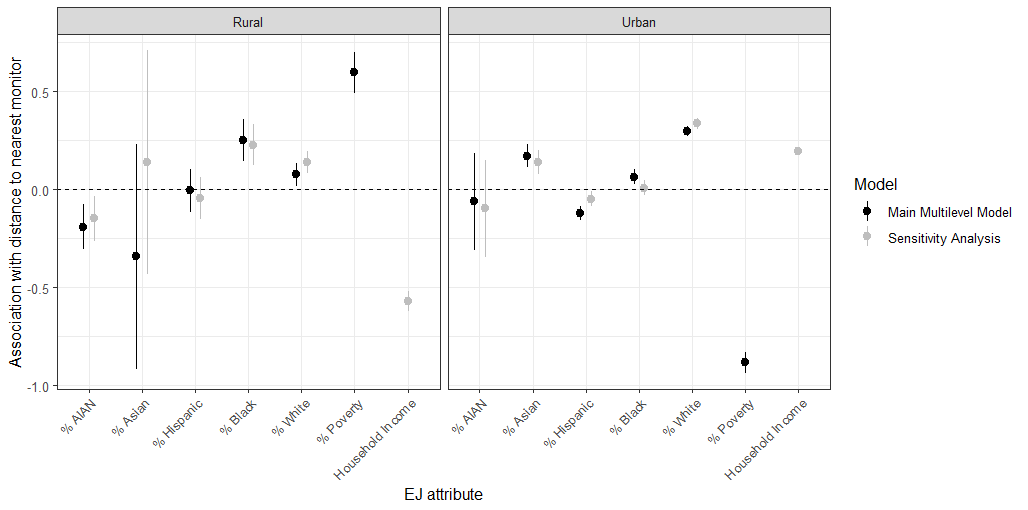}
\caption{The association between the EJ attribute and the distance to the nearest PM monitor for rural and urban tracts. A comparison can be seen between the estimate of the EJ coefficient in models with SES operationalized as \% poverty (black) as in the main models and median household income (grey) in the sensitivity analysis. Circles indicate point estimates and vertical bars the 95\% confidence intervals. Coefficients are very similar with largely overlapping 95\% confidence intervals. Note that a higher value of \% poverty indicates a tract with lower SES whereas a higher value of median household income indicates higher SES.}
\label{fig:sensitivity}
\end{figure}

\newpage

\subsection{Tables with regional model coefficient estimates}\label{tb:coeffs_region}

\begin{landscape}
\begin{table}[!htbp] \centering 

    \caption{EJ Attribute: \%AIAN (Rural)}  
  \setlength{\tabcolsep}{1pt} 
\begin{tabular}{@{\extracolsep{0pt}}lD{.}{.}{-3} D{.}{.}{-3} D{.}{.}{-3} D{.}{.}{-3} D{.}{.}{-3} D{.}{.}{-3} D{.}{.}{-3} D{.}{.}{-3} D{.}{.}{-3} D{.}{.}{-3} } 
\\[-1.8ex]\hline 
\hline \\[-1.8ex] 
 & \multicolumn{10}{c}{\textit{EPA Region}} \\ 

\\[-1.8ex] & \multicolumn{1}{c}{(1)} & \multicolumn{1}{c}{(2)} & \multicolumn{1}{c}{(3)} & \multicolumn{1}{c}{(4)} & \multicolumn{1}{c}{(5)} & \multicolumn{1}{c}{(6)} & \multicolumn{1}{c}{(7)} & \multicolumn{1}{c}{(8)} & \multicolumn{1}{c}{(9)} & \multicolumn{1}{c}{(10)}\\ 
\hline \\[-1.8ex] 
 Prop. Poverty & 2.070^{***} & 1.173^{***} & 0.688^{***} & 0.308^{***} & 0.957^{***} & 0.575^{***} & 0.482^{**} & 1.097^{***} & 0.624^{**} & 0.621^{**} \\ 
  & (0.409) & (0.274) & (0.256) & (0.072) & (0.155) & (0.127) & (0.199) & (0.419) & (0.307) & (0.282) \\ 
  Prop. White & 0.824 & 0.699^{***} & 0.502^{**} & -0.058 & 0.053 & 0.083 & -0.157 & 0.046 & -0.103 & 0.164 \\ 
  & (0.621) & (0.252) & (0.198) & (0.042) & (0.138) & (0.066) & (0.186) & (0.305) & (0.209) & (0.170) \\ 
  Prop. AIAN & -0.325 & 0.517 & 1.145 & -0.317^{**} & -1.121^{***} & 0.076 & -0.952^{*} & -0.017 & 0.107 & -0.284 \\ 
  & (1.314) & (0.375) & (4.024) & (0.158) & (0.225) & (0.131) & (0.575) & (0.336) & (0.288) & (0.199) \\ 
  Pop. Density & -12.536^{**} & -1.883 & -1.311 & -5.160^{***} & -2.083^{**} & -2.583^{*} & -3.733 & -1.589 & -4.223 & -7.938^{***} \\ 
  & (6.022) & (1.205) & (1.058) & (1.374) & (0.994) & (1.319) & (2.528) & (3.615) & (3.932) & (2.719) \\ 
  PM2.5 Z Score & -0.252^{**} & -0.144 & -0.264^{***} & -0.373^{***} & -0.317^{***} & -0.374^{***} & -0.491^{***} & -0.107 & -0.298^{***} & -0.121^{***} \\ 
  & (0.116) & (0.118) & (0.080) & (0.033) & (0.035) & (0.033) & (0.058) & (0.085) & (0.042) & (0.038) \\ 
  Intercept & 8.945^{***} & 9.349^{***} & 9.581^{***} & 10.482^{***} & 10.238^{***} & 10.549^{***} & 10.766^{***} & 10.650^{***} & 10.562^{***} & 10.569^{***} \\ 
  & (0.622) & (0.506) & (0.220) & (0.105) & (0.145) & (0.083) & (0.209) & (0.311) & (0.209) & (0.375) \\ 
 \hline \\[-1.8ex] 
Observations & \multicolumn{1}{c}{485} & \multicolumn{1}{c}{526} & \multicolumn{1}{c}{779} & \multicolumn{1}{c}{3,195} & \multicolumn{1}{c}{2,206} & \multicolumn{1}{c}{1,742} & \multicolumn{1}{c}{1,001} & \multicolumn{1}{c}{592} & \multicolumn{1}{c}{410} & \multicolumn{1}{c}{517} \\ 
 \hline \hline

  & \multicolumn{10}{r}{$^{*}$p$<$0.1; $^{**}$p$<$0.05; $^{***}$p$<$0.01} \\ 
\end{tabular} 
\end{table} 
\end{landscape}

\begin{landscape}
\begin{table}[!htbp] \centering 

    \caption{EJ Attribute: \%AIAN (Urban)} 

  \setlength{\tabcolsep}{1pt} 
\begin{tabular}{@{\extracolsep{0pt}}lD{.}{.}{-3} D{.}{.}{-3} D{.}{.}{-3} D{.}{.}{-3} D{.}{.}{-3} D{.}{.}{-3} D{.}{.}{-3} D{.}{.}{-3} D{.}{.}{-3} D{.}{.}{-3} } 
\\[-1.8ex]\hline 
\hline \\[-1.8ex] 
 & \multicolumn{10}{c}{\textit{EPA Region}} \\ 

\\[-1.8ex] & \multicolumn{1}{c}{(1)} & \multicolumn{1}{c}{(2)} & \multicolumn{1}{c}{(3)} & \multicolumn{1}{c}{(4)} & \multicolumn{1}{c}{(5)} & \multicolumn{1}{c}{(6)} & \multicolumn{1}{c}{(7)} & \multicolumn{1}{c}{(8)} & \multicolumn{1}{c}{(9)} & \multicolumn{1}{c}{(10)}\\ 
\hline \\[-1.8ex] 
 Prop. Poverty & -1.113^{***} & -0.685^{***} & -0.463^{***} & -0.638^{***} & -0.845^{***} & -0.512^{***} & -0.428^{***} & -1.408^{***} & -0.310^{**} & -1.133^{***} \\ 
  & (0.173) & (0.082) & (0.098) & (0.054) & (0.059) & (0.067) & (0.127) & (0.164) & (0.156) & (0.085) \\ 
  Prop. White & 0.896^{***} & 0.354^{***} & 0.477^{***} & 0.138^{***} & 0.518^{***} & 0.101^{***} & 0.329^{***} & -0.141 & 0.672^{***} & 0.134^{***} \\ 
  & (0.076) & (0.033) & (0.047) & (0.027) & (0.030) & (0.034) & (0.067) & (0.099) & (0.093) & (0.036) \\ 
  Prop. AIAN & 1.555^{*} & -0.313 & -0.067 & 0.264 & -1.555^{***} & 0.332 & -1.676 & 0.233 & -0.087 & -0.500^{**} \\ 
  & (0.858) & (0.651) & (1.296) & (0.335) & (0.600) & (0.229) & (1.150) & (0.608) & (0.481) & (0.242) \\ 
  Pop. Density & -0.519^{***} & -0.093^{***} & -0.313^{***} & -0.636^{***} & -0.243^{***} & -0.704^{***} & -1.974^{***} & -0.734^{***} & -0.322^{***} & -0.122^{***} \\ 
  & (0.055) & (0.010) & (0.039) & (0.038) & (0.027) & (0.053) & (0.142) & (0.123) & (0.062) & (0.025) \\ 
  PM2.5 Z Score & -0.888^{***} & -0.496^{***} & -0.579^{***} & -0.383^{***} & -0.593^{***} & -0.256^{***} & -0.607^{***} & -0.527^{***} & -0.776^{***} & -0.239^{***} \\ 
  & (0.056) & (0.027) & (0.026) & (0.016) & (0.025) & (0.018) & (0.050) & (0.049) & (0.034) & (0.009) \\ 
  Intercept & 7.854^{***} & 9.152^{***} & 9.143^{***} & 10.181^{***} & 9.655^{***} & 10.477^{***} & 10.108^{***} & 9.538^{***} & 9.421^{***} & 9.911^{***} \\ 
  & (0.178) & (0.430) & (0.165) & (0.086) & (0.132) & (0.094) & (0.237) & (0.214) & (0.354) & (0.410) \\ 
 \hline \\[-1.8ex] 
Observations & \multicolumn{1}{c}{3,151} & \multicolumn{1}{c}{6,839} & \multicolumn{1}{c}{5,057} & \multicolumn{1}{c}{13,792} & \multicolumn{1}{c}{11,746} & \multicolumn{1}{c}{8,985} & \multicolumn{1}{c}{2,900} & \multicolumn{1}{c}{2,487} & \multicolumn{1}{c}{2,793} & \multicolumn{1}{c}{10,975} \\ 

\hline 
\hline \\[-1.8ex] 
 & \multicolumn{10}{r}{$^{*}$p$<$0.1; $^{**}$p$<$0.05; $^{***}$p$<$0.01} \\ 
\end{tabular} 
\end{table} 
\end{landscape}

\begin{landscape}
\begin{table}[!htbp] \centering 

    \caption{EJ Attribute: \%Asian (Rural)} 
 
  \setlength{\tabcolsep}{1pt} 
\begin{tabular}{@{\extracolsep{0pt}}lD{.}{.}{-3} D{.}{.}{-3} D{.}{.}{-3} D{.}{.}{-3} D{.}{.}{-3} D{.}{.}{-3} D{.}{.}{-3} D{.}{.}{-3} D{.}{.}{-3} D{.}{.}{-3} } 
\\[-1.8ex]\hline 
\hline \\[-1.8ex] 
 & \multicolumn{10}{c}{\textit{EPA Region}} \\

\\[-1.8ex] & \multicolumn{1}{c}{(1)} & \multicolumn{1}{c}{(2)} & \multicolumn{1}{c}{(3)} & \multicolumn{1}{c}{(4)} & \multicolumn{1}{c}{(5)} & \multicolumn{1}{c}{(6)} & \multicolumn{1}{c}{(7)} & \multicolumn{1}{c}{(8)} & \multicolumn{1}{c}{(9)} & \multicolumn{1}{c}{(10)}\\ 
\hline \\[-1.8ex] 
 Prop. Poverty & 1.937^{***} & 1.210^{***} & 0.687^{***} & 0.311^{***} & 0.908^{***} & 0.589^{***} & 0.480^{**} & 1.076^{***} & 0.632^{**} & 0.525^{*} \\ 
  & (0.412) & (0.274) & (0.256) & (0.072) & (0.156) & (0.126) & (0.200) & (0.411) & (0.304) & (0.279) \\ 
  Prop. White & 0.455 & 0.566^{***} & 0.501^{**} & -0.044 & 0.488^{***} & 0.076 & -0.051 & 0.051 & -0.157 & 0.305^{**} \\ 
  & (0.582) & (0.170) & (0.202) & (0.041) & (0.107) & (0.063) & (0.178) & (0.174) & (0.156) & (0.138) \\ 
  Prop. Asian & -3.783^{*} & 2.717^{**} & 0.220 & -0.084 & -0.457 & 0.549 & -0.035 & -0.585 & -1.621 & -0.991 \\ 
  & (1.961) & (1.235) & (1.353) & (0.492) & (0.931) & (0.578) & (1.099) & (1.991) & (2.210) & (1.166) \\ 
  Pop. Density & -12.135^{**} & -2.386^{**} & -1.317 & -5.040^{***} & -1.502 & -2.686^{**} & -3.372 & -1.381 & -4.161 & -7.577^{***} \\ 
  & (5.962) & (1.204) & (1.058) & (1.375) & (0.996) & (1.312) & (2.519) & (3.648) & (3.927) & (2.714) \\ 
  PM2.5 Z Score & -0.239^{**} & -0.139 & -0.263^{***} & -0.375^{***} & -0.310^{***} & -0.376^{***} & -0.488^{***} & -0.106 & -0.300^{***} & -0.105^{***} \\ 
  & (0.116) & (0.117) & (0.080) & (0.033) & (0.035) & (0.033) & (0.058) & (0.085) & (0.041) & (0.036) \\ 
  Intercept & 9.348^{***} & 9.447^{***} & 9.582^{***} & 10.468^{***} & 9.832^{***} & 10.553^{***} & 10.663^{***} & 10.652^{***} & 10.618^{***} & 10.490^{***} \\ 
  & (0.598) & (0.480) & (0.224) & (0.105) & (0.122) & (0.082) & (0.203) & (0.225) & (0.178) & (0.362) \\ 
 \hline \\[-1.8ex] 
Observations & \multicolumn{1}{c}{485} & \multicolumn{1}{c}{526} & \multicolumn{1}{c}{779} & \multicolumn{1}{c}{3,195} & \multicolumn{1}{c}{2,206} & \multicolumn{1}{c}{1,742} & \multicolumn{1}{c}{1,001} & \multicolumn{1}{c}{592} & \multicolumn{1}{c}{410} & \multicolumn{1}{c}{517} \\ 

\hline 
\hline \\[-1.8ex] 
  & \multicolumn{10}{r}{$^{*}$p$<$0.1; $^{**}$p$<$0.05; $^{***}$p$<$0.01} \\ 
\end{tabular} 
\end{table} 
\end{landscape}

\begin{landscape}
\begin{table}[!htbp] \centering 

    \caption{EJ Attribute: \%Asian (Urban)} 

  \setlength{\tabcolsep}{1pt} 
\begin{tabular}{@{\extracolsep{0pt}}lD{.}{.}{-3} D{.}{.}{-3} D{.}{.}{-3} D{.}{.}{-3} D{.}{.}{-3} D{.}{.}{-3} D{.}{.}{-3} D{.}{.}{-3} D{.}{.}{-3} D{.}{.}{-3} } 
\\[-1.8ex]\hline 
\hline \\[-1.8ex] 
 & \multicolumn{10}{c}{\textit{EPA Region}} \\ 

\\[-1.8ex] & \multicolumn{1}{c}{(1)} & \multicolumn{1}{c}{(2)} & \multicolumn{1}{c}{(3)} & \multicolumn{1}{c}{(4)} & \multicolumn{1}{c}{(5)} & \multicolumn{1}{c}{(6)} & \multicolumn{1}{c}{(7)} & \multicolumn{1}{c}{(8)} & \multicolumn{1}{c}{(9)} & \multicolumn{1}{c}{(10)}\\ 
\hline \\[-1.8ex] 
 Prop. Poverty & -1.120^{***} & -0.703^{***} & -0.474^{***} & -0.619^{***} & -0.823^{***} & -0.469^{***} & -0.447^{***} & -1.429^{***} & -0.246 & -1.070^{***} \\ 
  & (0.173) & (0.082) & (0.099) & (0.054) & (0.059) & (0.068) & (0.127) & (0.162) & (0.157) & (0.086) \\ 
  Prop. White & 0.869^{***} & 0.342^{***} & 0.475^{***} & 0.138^{***} & 0.535^{***} & 0.102^{***} & 0.324^{***} & -0.087 & 0.805^{***} & 0.196^{***} \\ 
  & (0.075) & (0.034) & (0.047) & (0.027) & (0.029) & (0.034) & (0.066) & (0.098) & (0.104) & (0.037) \\ 
  Prop. Asian & -0.257 & -0.195^{***} & -0.228 & 0.348^{***} & 0.574^{***} & 0.432^{***} & -0.686^{**} & 1.884^{***} & 0.496^{***} & 0.328^{***} \\ 
  & (0.209) & (0.062) & (0.160) & (0.115) & (0.093) & (0.096) & (0.270) & (0.428) & (0.191) & (0.058) \\ 
  Pop. Density & -0.516^{***} & -0.094^{***} & -0.308^{***} & -0.641^{***} & -0.260^{***} & -0.725^{***} & -1.933^{***} & -0.746^{***} & -0.346^{***} & -0.119^{***} \\ 
  & (0.055) & (0.010) & (0.039) & (0.038) & (0.027) & (0.053) & (0.143) & (0.122) & (0.062) & (0.025) \\ 
  PM2.5 Z Score & -0.883^{***} & -0.507^{***} & -0.579^{***} & -0.388^{***} & -0.594^{***} & -0.264^{***} & -0.603^{***} & -0.534^{***} & -0.770^{***} & -0.234^{***} \\ 
  & (0.056) & (0.027) & (0.026) & (0.016) & (0.025) & (0.018) & (0.050) & (0.049) & (0.034) & (0.009) \\ 
  Intercept & 7.896^{***} & 9.163^{***} & 9.150^{***} & 10.173^{***} & 9.619^{***} & 10.474^{***} & 10.114^{***} & 9.468^{***} & 9.306^{***} & 9.838^{***} \\ 
  & (0.177) & (0.425) & (0.165) & (0.086) & (0.135) & (0.091) & (0.237) & (0.211) & (0.356) & (0.411) \\ 
 \hline \\[-1.8ex] 
Observations & \multicolumn{1}{c}{3,151} & \multicolumn{1}{c}{6,839} & \multicolumn{1}{c}{5,057} & \multicolumn{1}{c}{13,792} & \multicolumn{1}{c}{11,746} & \multicolumn{1}{c}{8,985} & \multicolumn{1}{c}{2,900} & \multicolumn{1}{c}{2,487} & \multicolumn{1}{c}{2,793} & \multicolumn{1}{c}{10,975} \\ 
 
\hline 
\hline \\[-1.8ex] 
  & \multicolumn{10}{r}{$^{*}$p$<$0.1; $^{**}$p$<$0.05; $^{***}$p$<$0.01} \\ 
\end{tabular} 
\end{table} 
\end{landscape}

\begin{landscape}
\begin{table}[!htbp] \centering 

    \caption{EJ Attribute: \%Black (Rural)} 

  \setlength{\tabcolsep}{1pt} 
\begin{tabular}{@{\extracolsep{0pt}}lD{.}{.}{-3} D{.}{.}{-3} D{.}{.}{-3} D{.}{.}{-3} D{.}{.}{-3} D{.}{.}{-3} D{.}{.}{-3} D{.}{.}{-3} D{.}{.}{-3} D{.}{.}{-3} } 
\\[-1.8ex]\hline 
\hline \\[-1.8ex] 
 & \multicolumn{10}{c}{\textit{EPA Region}} \\ 

\\[-1.8ex] & \multicolumn{1}{c}{(1)} & \multicolumn{1}{c}{(2)} & \multicolumn{1}{c}{(3)} & \multicolumn{1}{c}{(4)} & \multicolumn{1}{c}{(5)} & \multicolumn{1}{c}{(6)} & \multicolumn{1}{c}{(7)} & \multicolumn{1}{c}{(8)} & \multicolumn{1}{c}{(9)} & \multicolumn{1}{c}{(10)}\\ 
\hline \\[-1.8ex] 
 Prop. Poverty & 2.059^{***} & 1.152^{***} & 0.679^{***} & 0.300^{***} & 0.919^{***} & 0.557^{***} & 0.469^{**} & 1.086^{***} & 0.647^{**} & 0.572^{**} \\ 
  & (0.410) & (0.273) & (0.256) & (0.072) & (0.155) & (0.126) & (0.200) & (0.409) & (0.304) & (0.277) \\ 
  Prop. White & 0.933 & 0.349^{*} & 1.060^{***} & 0.340^{***} & 0.614^{***} & 0.184^{**} & 0.024 & 0.053 & -0.154 & 0.260^{*} \\ 
  & (0.580) & (0.180) & (0.365) & (0.082) & (0.120) & (0.074) & (0.203) & (0.175) & (0.156) & (0.140) \\ 
  Prop. Black & 0.305 & -0.488 & 0.993^{*} & 0.469^{***} & 0.563^{**} & 0.333^{***} & 0.259 & -0.176 & 1.410 & -1.232^{*} \\ 
  & (2.124) & (0.509) & (0.535) & (0.087) & (0.268) & (0.119) & (0.362) & (1.100) & (2.115) & (0.676) \\ 
  Pop. Density & -12.504^{**} & -2.103^{*} & -1.247 & -5.066^{***} & -1.573 & -2.786^{**} & -3.460 & -1.512 & -4.351 & -7.284^{***} \\ 
  & (6.056) & (1.201) & (1.057) & (1.369) & (0.993) & (1.310) & (2.522) & (3.608) & (3.918) & (2.710) \\ 
  PM2.5 Z Score & -0.252^{**} & -0.128 & -0.281^{***} & -0.360^{***} & -0.313^{***} & -0.380^{***} & -0.492^{***} & -0.106 & -0.301^{***} & -0.106^{***} \\ 
  & (0.116) & (0.117) & (0.081) & (0.033) & (0.035) & (0.033) & (0.058) & (0.085) & (0.041) & (0.036) \\ 
  Intercept & 8.839^{***} & 9.698^{***} & 9.014^{***} & 10.104^{***} & 9.704^{***} & 10.458^{***} & 10.590^{***} & 10.647^{***} & 10.594^{***} & 10.519^{***} \\ 
  & (0.592) & (0.480) & (0.379) & (0.122) & (0.133) & (0.091) & (0.223) & (0.225) & (0.180) & (0.368) \\ 
 \hline \\[-1.8ex] 
Observations & \multicolumn{1}{c}{485} & \multicolumn{1}{c}{526} & \multicolumn{1}{c}{779} & \multicolumn{1}{c}{3,195} & \multicolumn{1}{c}{2,206} & \multicolumn{1}{c}{1,742} & \multicolumn{1}{c}{1,001} & \multicolumn{1}{c}{592} & \multicolumn{1}{c}{410} & \multicolumn{1}{c}{517} \\ 
 
\hline 
\hline \\[-1.8ex] 
  & \multicolumn{10}{r}{$^{*}$p$<$0.1; $^{**}$p$<$0.05; $^{***}$p$<$0.01} \\ 
\end{tabular} 
\end{table} 
\end{landscape}

\begin{landscape}
\begin{table}[!htbp] \centering 

    \caption{EJ Attribute: \%Black (Urban)} 

  \setlength{\tabcolsep}{1pt} 
\begin{tabular}{@{\extracolsep{0pt}}lD{.}{.}{-3} D{.}{.}{-3} D{.}{.}{-3} D{.}{.}{-3} D{.}{.}{-3} D{.}{.}{-3} D{.}{.}{-3} D{.}{.}{-3} D{.}{.}{-3} D{.}{.}{-3} } 
\\[-1.8ex]\hline 
\hline \\[-1.8ex] 
 & \multicolumn{10}{c}{\textit{EPA Region}} \\ 

\\[-1.8ex] & \multicolumn{1}{c}{(1)} & \multicolumn{1}{c}{(2)} & \multicolumn{1}{c}{(3)} & \multicolumn{1}{c}{(4)} & \multicolumn{1}{c}{(5)} & \multicolumn{1}{c}{(6)} & \multicolumn{1}{c}{(7)} & \multicolumn{1}{c}{(8)} & \multicolumn{1}{c}{(9)} & \multicolumn{1}{c}{(10)}\\ 
\hline \\[-1.8ex] 
 Prop. Poverty & -1.158^{***} & -0.680^{***} & -0.459^{***} & -0.585^{***} & -0.939^{***} & -0.530^{***} & -0.441^{***} & -1.401^{***} & -0.221 & -1.134^{***} \\ 
  & (0.173) & (0.082) & (0.098) & (0.054) & (0.059) & (0.068) & (0.126) & (0.162) & (0.154) & (0.086) \\ 
  Prop. White & 0.600^{***} & 0.431^{***} & 0.729^{***} & -0.112^{***} & 0.913^{***} & 0.186^{***} & 0.971^{***} & 0.092 & 0.291^{***} & 0.137^{***} \\ 
  & (0.103) & (0.045) & (0.081) & (0.043) & (0.046) & (0.041) & (0.112) & (0.109) & (0.103) & (0.037) \\ 
  Prop. Black & -0.639^{***} & 0.117^{**} & 0.312^{***} & -0.324^{***} & 0.510^{***} & 0.185^{***} & 0.850^{***} & 1.512^{***} & -2.406^{***} & -0.098 \\ 
  & (0.164) & (0.047) & (0.081) & (0.044) & (0.046) & (0.050) & (0.121) & (0.319) & (0.302) & (0.095) \\ 
  Pop. Density & -0.552^{***} & -0.089^{***} & -0.277^{***} & -0.665^{***} & -0.195^{***} & -0.688^{***} & -1.779^{***} & -0.754^{***} & -0.349^{***} & -0.122^{***} \\ 
  & (0.056) & (0.010) & (0.040) & (0.038) & (0.027) & (0.053) & (0.143) & (0.122) & (0.061) & (0.025) \\ 
  PM2.5 Z Score & -0.893^{***} & -0.493^{***} & -0.569^{***} & -0.394^{***} & -0.563^{***} & -0.251^{***} & -0.601^{***} & -0.522^{***} & -0.761^{***} & -0.237^{***} \\ 
  & (0.056) & (0.027) & (0.026) & (0.016) & (0.025) & (0.018) & (0.049) & (0.049) & (0.033) & (0.009) \\ 
  Intercept & 8.134^{***} & 9.088^{***} & 8.906^{***} & 10.413^{***} & 9.303^{***} & 10.414^{***} & 9.537^{***} & 9.339^{***} & 9.741^{***} & 9.897^{***} \\ 
  & (0.187) & (0.427) & (0.177) & (0.094) & (0.131) & (0.103) & (0.240) & (0.216) & (0.355) & (0.405) \\ 
 \hline \\[-1.8ex] 
Observations & \multicolumn{1}{c}{3,151} & \multicolumn{1}{c}{6,839} & \multicolumn{1}{c}{5,057} & \multicolumn{1}{c}{13,792} & \multicolumn{1}{c}{11,746} & \multicolumn{1}{c}{8,985} & \multicolumn{1}{c}{2,900} & \multicolumn{1}{c}{2,487} & \multicolumn{1}{c}{2,793} & \multicolumn{1}{c}{10,975} \\ 

\hline 
\hline \\[-1.8ex] 
  & \multicolumn{10}{r}{$^{*}$p$<$0.1; $^{**}$p$<$0.05; $^{***}$p$<$0.01} \\ 
\end{tabular} 
\end{table} 
\end{landscape}

\begin{landscape}
\begin{table}[!htbp] \centering 

    \caption{EJ Attribute: \%Hispanic (Rural)} 

  \setlength{\tabcolsep}{1pt} 
\begin{tabular}{@{\extracolsep{0pt}}lD{.}{.}{-3} D{.}{.}{-3} D{.}{.}{-3} D{.}{.}{-3} D{.}{.}{-3} D{.}{.}{-3} D{.}{.}{-3} D{.}{.}{-3} D{.}{.}{-3} D{.}{.}{-3} } 
\\[-1.8ex]\hline 
\hline \\[-1.8ex] 
 & \multicolumn{10}{c}{\textit{EPA Region}} \\ 

\\[-1.8ex] & \multicolumn{1}{c}{(1)} & \multicolumn{1}{c}{(2)} & \multicolumn{1}{c}{(3)} & \multicolumn{1}{c}{(4)} & \multicolumn{1}{c}{(5)} & \multicolumn{1}{c}{(6)} & \multicolumn{1}{c}{(7)} & \multicolumn{1}{c}{(8)} & \multicolumn{1}{c}{(9)} & \multicolumn{1}{c}{(10)}\\ 
\hline \\[-1.8ex] 
 Prop. Poverty & 2.078^{***} & 1.184^{***} & 0.667^{***} & 0.307^{***} & 0.931^{***} & 0.542^{***} & 0.501^{**} & 1.097^{***} & 0.661^{**} & 0.671^{**} \\ 
  & (0.409) & (0.273) & (0.256) & (0.072) & (0.155) & (0.127) & (0.200) & (0.415) & (0.308) & (0.280) \\ 
  Prop. White & 1.065^{*} & 0.235 & 0.024 & -0.097^{**} & 0.780^{***} & -0.057 & 0.275 & 0.063 & -0.103 & 0.544^{***} \\ 
  & (0.645) & (0.180) & (0.260) & (0.043) & (0.125) & (0.082) & (0.244) & (0.190) & (0.211) & (0.166) \\ 
  Prop. Hispanic & 0.679 & -1.181^{**} & -1.516^{***} & -0.442^{***} & 1.055^{***} & -0.246^{**} & 0.661^{*} & 0.022 & 0.105 & 0.511^{**} \\ 
  & (1.494) & (0.520) & (0.571) & (0.102) & (0.245) & (0.098) & (0.347) & (0.369) & (0.283) & (0.200) \\ 
  Pop. Density & -12.527^{**} & -1.823 & -1.397 & -4.957^{***} & -2.034^{**} & -2.553^{*} & -3.366 & -1.583 & -4.467 & -8.151^{***} \\ 
  & (5.994) & (1.200) & (1.055) & (1.371) & (0.997) & (1.310) & (2.516) & (3.600) & (3.933) & (2.703) \\ 
  PM2.5 Z Score & -0.254^{**} & -0.163 & -0.276^{***} & -0.366^{***} & -0.315^{***} & -0.376^{***} & -0.478^{***} & -0.107 & -0.304^{***} & -0.138^{***} \\ 
  & (0.116) & (0.118) & (0.080) & (0.033) & (0.035) & (0.033) & (0.058) & (0.084) & (0.042) & (0.038) \\ 
  Intercept & 8.700^{***} & 9.805^{***} & 10.034^{***} & 10.531^{***} & 9.525^{***} & 10.690^{***} & 10.342^{***} & 10.633^{***} & 10.541^{***} & 10.168^{***} \\ 
  & (0.664) & (0.459) & (0.279) & (0.104) & (0.139) & (0.102) & (0.261) & (0.243) & (0.242) & (0.386) \\ 
 \hline \\[-1.8ex] 
Observations & \multicolumn{1}{c}{485} & \multicolumn{1}{c}{526} & \multicolumn{1}{c}{779} & \multicolumn{1}{c}{3,195} & \multicolumn{1}{c}{2,206} & \multicolumn{1}{c}{1,742} & \multicolumn{1}{c}{1,001} & \multicolumn{1}{c}{592} & \multicolumn{1}{c}{410} & \multicolumn{1}{c}{517} \\ 

\hline 
\hline \\[-1.8ex] 
 & \multicolumn{10}{r}{$^{*}$p$<$0.1; $^{**}$p$<$0.05; $^{***}$p$<$0.01} \\ 
\end{tabular} 
\end{table} 
\end{landscape}

\begin{landscape}
\begin{table}[!htbp] \centering 

    \caption{EJ Attribute: \%Hispanic (Urban)} 

  \setlength{\tabcolsep}{1pt} 
\begin{tabular}{@{\extracolsep{0pt}}lD{.}{.}{-3} D{.}{.}{-3} D{.}{.}{-3} D{.}{.}{-3} D{.}{.}{-3} D{.}{.}{-3} D{.}{.}{-3} D{.}{.}{-3} D{.}{.}{-3} D{.}{.}{-3} } 
\\[-1.8ex]\hline 
\hline \\[-1.8ex] 
 & \multicolumn{10}{c}{\textit{EPA Region}} \\ 

\\[-1.8ex] & \multicolumn{1}{c}{(1)} & \multicolumn{1}{c}{(2)} & \multicolumn{1}{c}{(3)} & \multicolumn{1}{c}{(4)} & \multicolumn{1}{c}{(5)} & \multicolumn{1}{c}{(6)} & \multicolumn{1}{c}{(7)} & \multicolumn{1}{c}{(8)} & \multicolumn{1}{c}{(9)} & \multicolumn{1}{c}{(10)}\\ 
\hline \\[-1.8ex] 
 Prop. Poverty & -1.182^{***} & -0.681^{***} & -0.447^{***} & -0.607^{***} & -0.955^{***} & -0.513^{***} & -0.440^{***} & -1.471^{***} & -0.357^{**} & -1.119^{***} \\ 
  & (0.174) & (0.082) & (0.098) & (0.054) & (0.059) & (0.067) & (0.126) & (0.161) & (0.156) & (0.085) \\ 
  Prop. White & 1.142^{***} & 0.347^{***} & 0.419^{***} & 0.204^{***} & 0.377^{***} & -0.041 & 0.165^{**} & -1.321^{***} & 1.008^{***} & -0.052 \\ 
  & (0.106) & (0.037) & (0.050) & (0.028) & (0.031) & (0.041) & (0.072) & (0.193) & (0.124) & (0.053) \\ 
  Prop. Hispanic & 0.517^{***} & -0.032 & -0.324^{***} & 0.301^{***} & -0.708^{***} & -0.268^{***} & -0.802^{***} & -1.557^{***} & 0.687^{***} & -0.255^{***} \\ 
  & (0.146) & (0.053) & (0.090) & (0.047) & (0.049) & (0.045) & (0.135) & (0.223) & (0.172) & (0.051) \\ 
  Pop. Density & -0.536^{***} & -0.092^{***} & -0.281^{***} & -0.659^{***} & -0.198^{***} & -0.692^{***} & -1.859^{***} & -0.774^{***} & -0.297^{***} & -0.117^{***} \\ 
  & (0.055) & (0.010) & (0.040) & (0.038) & (0.027) & (0.053) & (0.142) & (0.121) & (0.062) & (0.025) \\ 
  PM2.5 Z Score & -0.878^{***} & -0.493^{***} & -0.569^{***} & -0.389^{***} & -0.553^{***} & -0.254^{***} & -0.602^{***} & -0.535^{***} & -0.779^{***} & -0.235^{***} \\ 
  & (0.056) & (0.027) & (0.026) & (0.016) & (0.025) & (0.018) & (0.049) & (0.049) & (0.033) & (0.009) \\ 
  Intercept & 7.639^{***} & 9.162^{***} & 9.202^{***} & 10.112^{***} & 9.828^{***} & 10.621^{***} & 10.310^{***} & 10.671^{***} & 9.060^{***} & 10.076^{***} \\ 
  & (0.186) & (0.430) & (0.166) & (0.089) & (0.127) & (0.108) & (0.228) & (0.266) & (0.361) & (0.410) \\ 
 \hline \\[-1.8ex] 
Observations & \multicolumn{1}{c}{3,151} & \multicolumn{1}{c}{6,839} & \multicolumn{1}{c}{5,057} & \multicolumn{1}{c}{13,792} & \multicolumn{1}{c}{11,746} & \multicolumn{1}{c}{8,985} & \multicolumn{1}{c}{2,900} & \multicolumn{1}{c}{2,487} & \multicolumn{1}{c}{2,793} & \multicolumn{1}{c}{10,975} \\ 
 
\hline 
\hline \\[-1.8ex] 
  & \multicolumn{10}{r}{$^{*}$p$<$0.1; $^{**}$p$<$0.05; $^{***}$p$<$0.01} \\ 
\end{tabular} 
\end{table} 
\end{landscape}

\begin{landscape}
\begin{table}[!htbp] \centering 

    \caption{EJ Attribute: \%Poverty (Rural)} 

  \setlength{\tabcolsep}{1pt} 
\begin{tabular}{@{\extracolsep{0pt}}lD{.}{.}{-3} D{.}{.}{-3} D{.}{.}{-3} D{.}{.}{-3} D{.}{.}{-3} D{.}{.}{-3} D{.}{.}{-3} D{.}{.}{-3} D{.}{.}{-3} D{.}{.}{-3} } 
\\[-1.8ex]\hline 
\hline \\[-1.8ex] 
 & \multicolumn{10}{c}{\textit{EPA Region}} \\ 

\\[-1.8ex] & \multicolumn{1}{c}{(1)} & \multicolumn{1}{c}{(2)} & \multicolumn{1}{c}{(3)} & \multicolumn{1}{c}{(4)} & \multicolumn{1}{c}{(5)} & \multicolumn{1}{c}{(6)} & \multicolumn{1}{c}{(7)} & \multicolumn{1}{c}{(8)} & \multicolumn{1}{c}{(9)} & \multicolumn{1}{c}{(10)}\\ 
\hline \\[-1.8ex] 
 Prop. Poverty & 2.064^{***} & 1.140^{***} & 0.687^{***} & 0.311^{***} & 0.913^{***} & 0.585^{***} & 0.480^{**} & 1.092^{***} & 0.642^{**} & 0.549^{**} \\ 
  & (0.408) & (0.273) & (0.256) & (0.072) & (0.156) & (0.126) & (0.199) & (0.407) & (0.303) & (0.277) \\ 
  Prop. non-White & -0.902^{*} & -0.429^{***} & -0.493^{**} & 0.044 & -0.496^{***} & -0.073 & 0.050 & -0.058 & 0.155 & -0.306^{**} \\ 
  & (0.536) & (0.159) & (0.195) & (0.041) & (0.106) & (0.063) & (0.174) & (0.172) & (0.156) & (0.138) \\ 
  Pop. Density & -12.372^{**} & -2.051^{*} & -1.314 & -5.051^{***} & -1.529 & -2.662^{**} & -3.373 & -1.566 & -4.350 & -7.586^{***} \\ 
  & (5.979) & (1.199) & (1.057) & (1.373) & (0.994) & (1.312) & (2.517) & (3.586) & (3.914) & (2.712) \\ 
  PM2.5 Z Score & -0.251^{**} & -0.122 & -0.264^{***} & -0.375^{***} & -0.310^{***} & -0.375^{***} & -0.488^{***} & -0.107 & -0.301^{***} & -0.104^{***} \\ 
  & (0.116) & (0.117) & (0.080) & (0.033) & (0.035) & (0.032) & (0.058) & (0.084) & (0.041) & (0.036) \\ 
  Intercept & 9.773^{***} & 10.051^{***} & 10.084^{***} & 10.423^{***} & 10.318^{***} & 10.631^{***} & 10.612^{***} & 10.698^{***} & 10.447^{***} & 10.780^{***} \\ 
  & (0.198) & (0.455) & (0.113) & (0.099) & (0.062) & (0.068) & (0.113) & (0.158) & (0.121) & (0.356) \\ 
 \hline \\[-1.8ex] 
Observations & \multicolumn{1}{c}{485} & \multicolumn{1}{c}{526} & \multicolumn{1}{c}{779} & \multicolumn{1}{c}{3,195} & \multicolumn{1}{c}{2,206} & \multicolumn{1}{c}{1,742} & \multicolumn{1}{c}{1,001} & \multicolumn{1}{c}{592} & \multicolumn{1}{c}{410} & \multicolumn{1}{c}{517} \\ 

\hline 
\hline \\[-1.8ex] 
  & \multicolumn{10}{r}{$^{*}$p$<$0.1; $^{**}$p$<$0.05; $^{***}$p$<$0.01} \\ 
\end{tabular} 
\end{table} 
\end{landscape}

\begin{landscape}
\begin{table}[!htbp] \centering 

    \caption{EJ Attribute: \%Poverty (Urban)} 

  \setlength{\tabcolsep}{1pt} 
\begin{tabular}{@{\extracolsep{0pt}}lD{.}{.}{-3} D{.}{.}{-3} D{.}{.}{-3} D{.}{.}{-3} D{.}{.}{-3} D{.}{.}{-3} D{.}{.}{-3} D{.}{.}{-3} D{.}{.}{-3} D{.}{.}{-3} } 
\\[-1.8ex]\hline 
\hline \\[-1.8ex] 
 & \multicolumn{10}{c}{\textit{EPA Region}} \\ 

\\[-1.8ex] & \multicolumn{1}{c}{(1)} & \multicolumn{1}{c}{(2)} & \multicolumn{1}{c}{(3)} & \multicolumn{1}{c}{(4)} & \multicolumn{1}{c}{(5)} & \multicolumn{1}{c}{(6)} & \multicolumn{1}{c}{(7)} & \multicolumn{1}{c}{(8)} & \multicolumn{1}{c}{(9)} & \multicolumn{1}{c}{(10)}\\ 
\hline \\[-1.8ex] 
 Prop. Poverty & -1.113^{***} & -0.686^{***} & -0.463^{***} & -0.638^{***} & -0.847^{***} & -0.509^{***} & -0.437^{***} & -1.400^{***} & -0.311^{**} & -1.145^{***} \\ 
  & (0.173) & (0.082) & (0.098) & (0.054) & (0.059) & (0.067) & (0.127) & (0.163) & (0.156) & (0.085) \\ 
  Prop. non-White & -0.879^{***} & -0.356^{***} & -0.477^{***} & -0.137^{***} & -0.524^{***} & -0.098^{***} & -0.337^{***} & 0.149 & -0.676^{***} & -0.146^{***} \\ 
  & (0.075) & (0.033) & (0.047) & (0.027) & (0.029) & (0.034) & (0.066) & (0.097) & (0.091) & (0.036) \\ 
  Pop. Density & -0.522^{***} & -0.093^{***} & -0.313^{***} & -0.636^{***} & -0.245^{***} & -0.706^{***} & -1.977^{***} & -0.737^{***} & -0.322^{***} & -0.121^{***} \\ 
  & (0.055) & (0.010) & (0.039) & (0.038) & (0.027) & (0.053) & (0.142) & (0.122) & (0.062) & (0.025) \\ 
  PM2.5 Z Score & -0.889^{***} & -0.496^{***} & -0.579^{***} & -0.383^{***} & -0.594^{***} & -0.256^{***} & -0.607^{***} & -0.527^{***} & -0.776^{***} & -0.238^{***} \\ 
  & (0.056) & (0.027) & (0.026) & (0.016) & (0.025) & (0.018) & (0.050) & (0.049) & (0.034) & (0.009) \\ 
  Intercept & 8.753^{***} & 9.506^{***} & 9.620^{***} & 10.320^{***} & 10.164^{***} & 10.588^{***} & 10.428^{***} & 9.401^{***} & 10.093^{***} & 10.036^{***} \\ 
  & (0.163) & (0.429) & (0.159) & (0.083) & (0.132) & (0.090) & (0.231) & (0.199) & (0.346) & (0.405) \\ 
 \hline \\[-1.8ex] 
Observations & \multicolumn{1}{c}{3,151} & \multicolumn{1}{c}{6,839} & \multicolumn{1}{c}{5,057} & \multicolumn{1}{c}{13,792} & \multicolumn{1}{c}{11,746} & \multicolumn{1}{c}{8,985} & \multicolumn{1}{c}{2,900} & \multicolumn{1}{c}{2,487} & \multicolumn{1}{c}{2,793} & \multicolumn{1}{c}{10,975} \\ 

\hline 
\hline \\[-1.8ex] 
  & \multicolumn{10}{r}{$^{*}$p$<$0.1; $^{**}$p$<$0.05; $^{***}$p$<$0.01} \\ 
\end{tabular} 
\end{table} 
\end{landscape}

\begin{landscape}
\begin{table}[!htbp] \centering 

    \caption{EJ Attribute: \%White (Rural)} 

  \setlength{\tabcolsep}{1pt} 
\begin{tabular}{@{\extracolsep{0pt}}lD{.}{.}{-3} D{.}{.}{-3} D{.}{.}{-3} D{.}{.}{-3} D{.}{.}{-3} D{.}{.}{-3} D{.}{.}{-3} D{.}{.}{-3} D{.}{.}{-3} D{.}{.}{-3} } 
\\[-1.8ex]\hline 
\hline \\[-1.8ex] 
 & \multicolumn{10}{c}{\textit{EPA Region}} \\ 

\\[-1.8ex] & \multicolumn{1}{c}{(1)} & \multicolumn{1}{c}{(2)} & \multicolumn{1}{c}{(3)} & \multicolumn{1}{c}{(4)} & \multicolumn{1}{c}{(5)} & \multicolumn{1}{c}{(6)} & \multicolumn{1}{c}{(7)} & \multicolumn{1}{c}{(8)} & \multicolumn{1}{c}{(9)} & \multicolumn{1}{c}{(10)}\\ 
\hline \\[-1.8ex] 
 Prop. Poverty & 2.064^{***} & 1.140^{***} & 0.687^{***} & 0.311^{***} & 0.913^{***} & 0.585^{***} & 0.480^{**} & 1.092^{***} & 0.642^{**} & 0.549^{**} \\ 
  & (0.408) & (0.273) & (0.256) & (0.072) & (0.156) & (0.126) & (0.199) & (0.407) & (0.303) & (0.277) \\ 
  Prop. White & 0.902^{*} & 0.429^{***} & 0.493^{**} & -0.044 & 0.496^{***} & 0.073 & -0.050 & 0.058 & -0.155 & 0.306^{**} \\ 
  & (0.536) & (0.159) & (0.195) & (0.041) & (0.106) & (0.063) & (0.174) & (0.172) & (0.156) & (0.138) \\ 
  Pop. Density & -12.372^{**} & -2.051^{*} & -1.314 & -5.051^{***} & -1.529 & -2.662^{**} & -3.373 & -1.566 & -4.350 & -7.586^{***} \\ 
  & (5.979) & (1.199) & (1.057) & (1.373) & (0.994) & (1.312) & (2.517) & (3.586) & (3.914) & (2.712) \\ 
  PM2.5 Z Score & -0.251^{**} & -0.122 & -0.264^{***} & -0.375^{***} & -0.310^{***} & -0.375^{***} & -0.488^{***} & -0.107 & -0.301^{***} & -0.104^{***} \\ 
  & (0.116) & (0.117) & (0.080) & (0.033) & (0.035) & (0.032) & (0.058) & (0.084) & (0.041) & (0.036) \\ 
  Intercept & 8.872^{***} & 9.623^{***} & 9.592^{***} & 10.467^{***} & 9.822^{***} & 10.558^{***} & 10.663^{***} & 10.640^{***} & 10.602^{***} & 10.474^{***} \\ 
  & (0.546) & (0.476) & (0.217) & (0.105) & (0.121) & (0.081) & (0.199) & (0.221) & (0.178) & (0.366) \\ 
 \hline \\[-1.8ex] 
Observations & \multicolumn{1}{c}{485} & \multicolumn{1}{c}{526} & \multicolumn{1}{c}{779} & \multicolumn{1}{c}{3,195} & \multicolumn{1}{c}{2,206} & \multicolumn{1}{c}{1,742} & \multicolumn{1}{c}{1,001} & \multicolumn{1}{c}{592} & \multicolumn{1}{c}{410} & \multicolumn{1}{c}{517} \\ 

\hline 
\hline \\[-1.8ex] 
  & \multicolumn{10}{r}{$^{*}$p$<$0.1; $^{**}$p$<$0.05; $^{***}$p$<$0.01} \\ 
\end{tabular} 
\end{table} 
\end{landscape}

\begin{landscape}
\begin{table}[!htbp] \centering 

    \caption{EJ Attribute: \%White (Urban)} 

  \setlength{\tabcolsep}{1pt} 
\begin{tabular}{@{\extracolsep{0pt}}lD{.}{.}{-3} D{.}{.}{-3} D{.}{.}{-3} D{.}{.}{-3} D{.}{.}{-3} D{.}{.}{-3} D{.}{.}{-3} D{.}{.}{-3} D{.}{.}{-3} D{.}{.}{-3} } 
\\[-1.8ex]\hline 
\hline \\[-1.8ex] 
 & \multicolumn{10}{c}{\textit{EPA Region}} \\ 
\cline{2-11} 
\\[-1.8ex] & \multicolumn{10}{c}{log\_dist} \\ 
\\[-1.8ex] & \multicolumn{1}{c}{(1)} & \multicolumn{1}{c}{(2)} & \multicolumn{1}{c}{(3)} & \multicolumn{1}{c}{(4)} & \multicolumn{1}{c}{(5)} & \multicolumn{1}{c}{(6)} & \multicolumn{1}{c}{(7)} & \multicolumn{1}{c}{(8)} & \multicolumn{1}{c}{(9)} & \multicolumn{1}{c}{(10)}\\ 
\hline \\[-1.8ex] 
 Prop. Poverty & -1.113^{***} & -0.686^{***} & -0.463^{***} & -0.638^{***} & -0.847^{***} & -0.509^{***} & -0.437^{***} & -1.400^{***} & -0.311^{**} & -1.145^{***} \\ 
  & (0.173) & (0.082) & (0.098) & (0.054) & (0.059) & (0.067) & (0.127) & (0.163) & (0.156) & (0.085) \\ 
  Prop. non-White & 0.879^{***} & 0.356^{***} & 0.477^{***} & 0.137^{***} & 0.524^{***} & 0.098^{***} & 0.337^{***} & -0.149 & 0.676^{***} & 0.146^{***} \\ 
  & (0.075) & (0.033) & (0.047) & (0.027) & (0.029) & (0.034) & (0.066) & (0.097) & (0.091) & (0.036) \\ 
  Pop. Density & -0.522^{***} & -0.093^{***} & -0.313^{***} & -0.636^{***} & -0.245^{***} & -0.706^{***} & -1.977^{***} & -0.737^{***} & -0.322^{***} & -0.121^{***} \\ 
  & (0.055) & (0.010) & (0.039) & (0.038) & (0.027) & (0.053) & (0.142) & (0.122) & (0.062) & (0.025) \\ 
  PM2.5 Z Score & -0.889^{***} & -0.496^{***} & -0.579^{***} & -0.383^{***} & -0.594^{***} & -0.256^{***} & -0.607^{***} & -0.527^{***} & -0.776^{***} & -0.238^{***} \\ 
  & (0.056) & (0.027) & (0.026) & (0.016) & (0.025) & (0.018) & (0.050) & (0.049) & (0.034) & (0.009) \\ 
  Intercept & 7.874^{***} & 9.150^{***} & 9.143^{***} & 10.182^{***} & 9.640^{***} & 10.490^{***} & 10.092^{***} & 9.550^{***} & 9.418^{***} & 9.890^{***} \\ 
  & (0.176) & (0.429) & (0.164) & (0.086) & (0.135) & (0.092) & (0.238) & (0.211) & (0.353) & (0.405) \\ 
 \hline \\[-1.8ex] 
Observations & \multicolumn{1}{c}{3,151} & \multicolumn{1}{c}{6,839} & \multicolumn{1}{c}{5,057} & \multicolumn{1}{c}{13,792} & \multicolumn{1}{c}{11,746} & \multicolumn{1}{c}{8,985} & \multicolumn{1}{c}{2,900} & \multicolumn{1}{c}{2,487} & \multicolumn{1}{c}{2,793} & \multicolumn{1}{c}{10,975} \\ 

\hline 
\hline \\[-1.8ex] 
  & \multicolumn{10}{r}{$^{*}$p$<$0.1; $^{**}$p$<$0.05; $^{***}$p$<$0.01} \\ 
\end{tabular} 
\end{table} 
\end{landscape}


\end{document}